\documentclass{article}
\pdfoutput=1
\usepackage[top=2cm,bottom=3cm,left=2cm,right=2cm]{geometry}
\usepackage{amsmath}
\usepackage{amsfonts}
\usepackage{hyperref}
\usepackage{graphicx}

\numberwithin{equation}{section}
\numberwithin{figure}{section}

\def\eq#1{(\ref{eq:#1})}
\def\lineup{\!\!\!\!\!\!\!\!\!\!&&}

\def\d{\partial}
\def\eps{\epsilon}
\def\Vspace{\phantom{\bigg(}}

\def\PhiR{\Phi_\mathrm{R}}
\def\PhiN{\Phi_\mathrm{NS}}

\def\H{\mathcal{H}}
\def\Hf{\mathcal{H}_{\Phi}}

\begin{document}

\begin{titlepage}
\rightline{\tt LMU-ASC 18/17}
\rightline\today

\begin{center}
\vskip 3.5cm

{\large \bf{One Loop Tadpole in Heterotic String Field Theory}}

\vskip 1.0cm

{\large {Theodore Erler$^{(b)}$\footnote{tchovi@gmail.com}, Sebastian Konopka$^{(a)}$\footnote{konopka.seb@googlemail.com}}, Ivo Sachs$^{(a)}$\footnote{ivo.sachs@physik.uni-muenchen.de}}

\vskip 1.0cm

$^{(a)}${\it {Arnold Sommerfeld Center, Ludwig-Maximilians University}}\\
{\it {Theresienstrasse 37, 80333 Munich, Germany}}\\

\vskip .5cm

$^{(b)}${\it Institute of Physics of the ASCR, v.v.i.}\\
{\it Na Slovance 2, 182 21 Prague 8, Czech Republic}\\

\vspace{.5cm}

\vskip 2.0cm

{\bf Abstract} 
\end{center}
We compute the off-shell 1-loop tadpole amplitude in heterotic string field theory. With a special choice of cubic vertex, we show that this amplitude can be computed exactly.  We obtain explicit and elementary expressions for the Feynman graph decomposition of the moduli space, the local coordinate map at the puncture as a function of the modulus, and the $b$-ghost insertions needed for the integration measure. Recently developed homotopy algebra methods provide a consistent configuration of picture changing operators. We discuss the consequences of spurious poles for the choice of picture changing operators.

\end{titlepage}

\tableofcontents

\section{Introduction}

In this paper we compute the off-shell, one-loop tadpole amplitude in heterotic string field theory. The purpose is twofold:
\begin{description}
\item{(1)} First, we would like to show that the amplitude can be computed exactly. Our success in this regard is largely due to a nonstandard choice of cubic vertex defined by $SL(2,\mathbb{C})$ local coordinate maps.\footnote{We thank A. Sen for suggesting the utility of $SL(2,\mathbb{C})$ maps. The cubic vertex we use was also discussed in \cite{Sonoda,Zwconst,SonodaZwiebach}. In particular \cite{Zwconst} describes the propagator contribution to the tadpole amplitude in closed bosonic string field theory using this cubic vertex. We thank B. Zwiebach for pointing us to these references.} We will take special care to provide explicit results concerning the Feynman-graph decomposition of the moduli space, the local coordinate map as a function of the modulus, and the $b$-ghost insertions needed for the integration measure. Actually, these data are primarily associated with closed bosonic string field theory, but they also represent the most significant obstacle to explicit results for the heterotic string.  Importantly, our string vertices differ from the canonical ones defined by minimal area metrics \cite{Zwiebach}. The minimal area vertices are cumbersome for elementary calculations, though some analytic results are available at tree level up to 4 points \cite{Belopolsky} and numerical calculations have been performed up to 5 points \cite{Moeller1,Moeller2}.\footnote{A new approach to the computation of off-shell amplitudes has recently been proposed based on hyperbolic geometry \cite{Pius}. It would be interesting to approach the computation of the heterotic tadpole from this perspective.}
\item{(2)} Second, we would like to see how recent homotopy-algebraic constructions of classical superstring field theories \cite{WittenSS,ClosedSS,Ramond,complete,RWaction,KSR,hetR} may be extended to the quantum level. A significant issue is the appearance of spurious poles in $\beta\gamma$ correlation functions at higher genus \cite{Verlinde}. We find that the general methodology behind tree level constructions still functions in loops. Spurious poles are mainly important for the choice of picture changing operators (PCOs), which must ensure that loop vertices are finite and unambiguously defined. We consider two approaches to inserting PCOs. In the first approach, PCOs appear at specific points in the surfaces defining the vertices, in a manner similar to \cite{SenWitten}. In the second approach, PCOs appear as contour integrals around the punctures, paralleling the construction of classical superstring field theories. The second approach has a somewhat different nature in loops, however, due to the necessity of specifying the PCO contours precisely in relation spurious poles. A naive treatment can lead to inconsistencies. 
\end{description}
In this paper we are not concerned with computing the 1-loop tadpole in a specific background or for any particular on- or off-shell states. So when we claim to ``compute" this amplitude, really what we mean is that we specify all background and state-independent string field theory data that goes into the definition of this amplitude. It will remain to choose a vertex operator of interest, compute the correlation functions, and integrate over the moduli space to obtain a final expression. Ultimately, however, one would like to use string field theory to compute physical amplitudes in situations where the conventional formulation of superstring perturbation theory breaks down. The computations of this paper can be regarded as a modest step in this direction. 

The paper is organized as follows. In section \ref{sec:closedSFT} we briefly review the definition of bosonic and heterotic closed string field theories. To obtain the data associated to integration over the bosonic moduli space, in section \ref{sec:bosonictad} we compute the 1-loop tadpole in closed bosonic string field theory. This requires, in particular, choosing a suitable cubic vertex, computing the contribution to the tadpole from gluing two legs of the cubic vertex with a propagator, and defining an appropriate fundamental tadpole vertex to fill in the remaining region of the moduli space. In section \ref{sec:hettad} we compute the 1-loop tadpole in heterotic string field theory. This requires dressing the amplitude of the bosonic string with a configuration of PCOs. Homotopy algebraic methods constrain the choice of PCOs to be consistent with quantum gauge invariance, but some freedom remains. We discuss two approaches: one which inserts PCOs at specific points in the surfaces defining the vertices, and another which inserts PCOs in the form of contour integrals around the punctures. We discuss the consequences of spurious poles for both approaches. We also present the amplitudes in a form which is manifestly well-defined in the small Hilbert space. We end with concluding remarks.



\section{Quantum Closed String Field Theories}
\label{sec:closedSFT}

In this section we review the definition of quantum closed string field theories, both bosonic and heterotic versions. For the closed bosonic string, we will only sketch the basic structure of the theory as suits our needs. Everything is taken from Zwiebach's seminal paper \cite{Zwiebach}, with some minor changes of notation to make contact with more recent work. The structure of heterotic string field theory closely parallels that of the closed bosonic string. We follow \cite{1PIR,SenBV} in our treatment of the Ramond sector. 

\subsection{Closed Bosonic String Field Theory}

Closed bosonic string field theory is the field theory of fluctuations of a closed string background in bosonic string theory. The background is described by a worldsheet conformal field theory, which factorizes into a $c=26$ matter component and a $c=-26$ $bc$ ghost component. A string field is an element of the state space $\H$ of this conformal field theory. The state space has two important gradings: ghost number, and Grassmann parity, which distinguishes commuting and anticommuting states. Since the field theory path integral requires anticommuting ghosts in the target space, we allow states in $\H$ to appear in linear combinations with Grassmann even or Grassmann odd coefficients.

Quantum closed bosonic string field theory is defined in the framework of the Batalin-Vilkovisky formalism, which gives a prescription for computing amplitudes via Feynman rules derived from a gauge-fixed path integral. The central object in the formalism is the quantum master action (henceforth simply the ``action"), which takes the form\footnote{We set $\hbar$ and the closed string coupling constant to $1$.}
\begin{eqnarray}
S \lineup = \frac{1}{2!}\omega(\Phi,Q\Phi) + \frac{1}{3!}\omega(\Phi,\ell_{0,2}(\Phi,\Phi)) + \frac{1}{4!}\omega(\Phi,\ell_{0,3}(\Phi,\Phi,\Phi)) + \frac{1}{5!}\omega(\Phi,\ell_{0,4}(\Phi,\Phi,\Phi,\Phi))+...\nonumber\\
\lineup\ \ \ \ \ \ \ \ \ \ \ \ \ \ \ \ \ \ \ \, +\,  \ \ \ \ \frac{1}{1!} \omega(\Phi,\ell_{1,0}) \ \ \ \  \,+\ \ \ \ \frac{1}{2!}\omega(\Phi,\ell_{1,1}(\Phi))\ \ \ +\ \ \ \ \frac{1}{3!}\omega(\Phi,\ell_{1,2}(\Phi,\Phi))\ \ \, +...\nonumber\\
\lineup\ \ \ \ \ \ \ \ \ \ \ \ \ \ \ \ \ \ \ \ \ \ \ \ \ \ \ \ \ \ \ \ \ \ \ \ \ \ \ \ \ \ \ \ \ \ \ \ \ \ \ \ \ \ \ \ \ \ \ \ \ \ \ \ \ \ \ \ \ \ \ \ \ \ \ \ \ \, +\ \ \ \ \ \ \ \ \frac{1}{1!}\omega(\Phi,\ell_{2,0})\ \ \ \ \ \ \ +...\ .\label{eq:bosmasteraction}
\end{eqnarray}
Let us describe the ingredients:

\vspace{.5cm}

\noindent {\it Dynamical string field $\Phi$}: The state $\Phi$ is the dynamical string field. It is not an arbitrary state in $\H$, but a Grassmann even element of a linear subspace
\begin{equation}\Hf\subset\H\end{equation}
characterized by states satisfying the $b_0^-$ and level matching constraints:
\begin{equation}b_0^- \Phi = L_0^-\Phi = 0,\end{equation}
where 
\begin{eqnarray}
b_0^- \lineup \equiv b_0 -\overline{b}_0\\
L_0^- \lineup \equiv L_0 - \overline{L}_0\\
b_0^+ \lineup \equiv b_0 +\overline{b}_0\\
L_0^+ \lineup \equiv L_0 + \overline{L}_0,
\end{eqnarray}
and the bar indicates operators from the antiholomorphic sector. Classically, the dynamical field $\Phi$ carries ghost number 2, but quantum mechanically it contains components at all ghost numbers which play the role of fields and antifields in the Batalin-Vilkovisky formalism. All states in $\Hf$ are proportional to the operator
\begin{equation}b_0^- \delta(L_0^-) \equiv b_0^- \int_{-\pi}^\pi \frac{d\theta}{2\pi} e^{i\theta L_0^-}.\end{equation}
Since $L_0^-$ takes only integer eigenvalues, one can see that
\begin{eqnarray}
L_0^- \delta(L_0^-)\lineup =\int_{-\pi}^\pi\frac{d\theta}{2\pi}L_0^- e^{i\theta L_0^-}\nonumber\\
\lineup = -i\int_{-\pi}^\pi\frac{d\theta}{2\pi}\frac{\d}{\d\theta} e^{i\theta L_0^-}\nonumber\\
\lineup = -i(e^{i\pi L_0^-}-e^{-i\pi L_0^-})\nonumber\\
\lineup = 0,\label{eq:LmdLm}
\end{eqnarray}
which motivates the delta function notation. Note that $\delta(L_0^-)$ is a projector, and $b_0^-\delta(L_0^-)$ is BRST invariant. The statement that $\Phi$ satisfies $b_0^-$ and level matching constraints is equivalent to
\begin{equation}
\Phi = b_0^- \delta(L_0^-)c_0^- \Phi,
\end{equation}
where 
\begin{equation}c_0^-\equiv\frac{1}{2}(c_0 -\overline{c}_0),\end{equation}
which satisfies $[b_0^-,c_0^-] = 1$.\footnote{Commutators are graded with respect to Grassmann parity.}

\vspace{.5cm}

\noindent{\it Symplectic form $\omega$}: The object $\omega$ is a symplectic form. It is a bilinear map from two copies of $\H_\Phi$ into numbers which is graded antisymmetric,
\begin{equation}\omega(A,B) = -(-1)^{AB}\omega(B,A),\end{equation}
and nondegenerate. The symplectic form is concretely defined 
\begin{equation}\omega(A,B) = (-1)^A \langle A,c_0^- B\rangle,\label{eq:bos_om}\end{equation}
where the bracket $\langle\cdot,\cdot\rangle$ denotes the BPZ inner product, which can be computed as a correlation function on the complex plane
\begin{equation}
\langle A,B\rangle \equiv \langle I\circ A(0) B(0)\rangle,
\end{equation}
where $A(0),B(0)$ are the vertex operators corresponding to the states $A,B$ and $I\circ$ denotes conformal transformation with the map $I(z)=1/z$. Often it is convenient to write the symplectic form and BPZ inner product as ``double bra" states:
\begin{eqnarray}
\langle\omega|:\lineup \Hf^{\otimes 2}\to \H^{\otimes 0}\\
\langle \mathrm{bpz}|:\lineup \H^{\otimes 2}\to \H^{\otimes 0},
\end{eqnarray}
so that 
\begin{eqnarray}
\langle \omega|A\otimes B \lineup = \omega(A,B)\\
\langle\mathrm{bpz}|A\otimes B\lineup = \langle A,B\rangle.
\end{eqnarray}
Note that the symplectic form is only defined operating on states in the subspace $\Hf$, whereas the BPZ inner product is defined operating on any states in $\H$. Since the operator $c_0^-$ is BPZ odd, we have the relation
\begin{equation}\langle \mathrm{bpz}|(c_0^-\otimes \mathbb{I} + \mathbb{I}\otimes c_0^-) = 0\end{equation}
where $\mathbb{I}$ is the identity operator on the state space. We may then relate the symplectic form and BPZ inner product with the formula
\begin{equation}
\langle \omega| = -\langle\mathrm{bpz}|c_0^-\otimes \mathbb{I}.
\end{equation}
Particularly important in the quantum theory is the inverse of the symplectic form, a Poisson bivector. We will write the Poisson bivector as a Grassmann odd ``double ket" state
\begin{equation} |\omega^{-1}\rangle \in \Hf^{\otimes 2},\end{equation}
which by definition satisfies\footnote{This can also be written as \begin{equation}\langle \omega|^{(12)}\, ^{(23)}|\omega^{-1}\rangle =\, ^{(3)}\mathbb{I}^{(1)},\ \ \ \langle\omega|^{(23)}\,^{(12)}|\omega^{-1}\rangle = \,^{(1)}\mathbb{I}^{(3)}\end{equation} where the superscripts denote copies $1,2,3$ of the state space. Pre-superscripts denote the output, and post-superscripts the input.}
\begin{equation}\big(\langle\omega| \otimes\mathbb{I}\big)\big(\mathbb{I}\otimes|\omega^{-1}\rangle\big) = \big(\mathbb{I}\otimes\langle\omega|\big)\big(|\omega^{-1}\rangle\otimes\mathbb{I}\big)=\mathbb{I}.\end{equation}
The Poisson bivector can be constructed as follows. Since the BPZ inner product is nondegenerate, there is a Grassmann even ``double ket" state
\begin{equation}
|\mathrm{bpz}^{-1}\rangle \in \H^{\otimes 2}
\end{equation}
satisfying
\begin{equation}\big(\langle\mathrm{bpz}| \otimes\mathbb{I}\big)\big(\mathbb{I}\otimes|\mathrm{bpz}^{-1}\rangle\big) = \big(\mathbb{I}\otimes\langle\mathrm{bpz}|\big)\big(|\mathrm{bpz}^{-1}\rangle\otimes\mathbb{I}\big)=\mathbb{I}.\end{equation}
Since $b_0^-$ and $L_0^-$ are BPZ even, it is straightforward to show that 
\begin{eqnarray}
(b_0^-\otimes\mathbb{I})|\mathrm{bpz}^{-1}\rangle \lineup = (\mathbb{I}\otimes b_0^-)|\mathrm{bpz}^{-1}\rangle\label{eq:b0bpzinv}\\
(L_0^-\otimes\mathbb{I})|\mathrm{bpz}^{-1}\rangle \lineup = (\mathbb{I}\otimes L_0^-)|\mathrm{bpz}^{-1}\rangle.\label{eq:L0bpzinv}
\end{eqnarray}
The Poisson bivector must satisfy $b_0^-$ and level matching constraints, so it is natural to guess the expression: 
\begin{equation}|\omega^{-1}\rangle = \big(b_0^-\delta(L_0^-)\otimes \mathbb{I}\big)|\mathrm{bpz}^{-1}\rangle.\label{eq:bosPois}\end{equation}
To check that this expression is correct, compute 
\begin{eqnarray}
\big(\langle \omega|\otimes\mathbb{I}\big)\big(\mathbb{I}\otimes |\omega^{-1}\rangle\big) \lineup = \big(\langle \mathrm{bpz}|\otimes\mathbb{I}\big)\big(-c_0^-\otimes\mathbb{I}\otimes\mathbb{I}\big)\big(\mathbb{I}\otimes b_0^-\delta(L_0^-)\otimes\mathbb{I}\big)\big(\mathbb{I}\otimes |\mathrm{bpz}^{-1}\rangle\big)\nonumber\\
\lineup = \big(\langle \mathrm{bpz}|\otimes\mathbb{I}\big)\big(\mathbb{I}\otimes b_0^-\delta(L_0^-)\otimes\mathbb{I}\big)\big(c_0^-\otimes\mathbb{I}\otimes\mathbb{I}\big)\big(\mathbb{I}\otimes |\mathrm{bpz}^{-1}\rangle\big)\nonumber\\
\lineup = \big(\langle \mathrm{bpz}|\otimes\mathbb{I}\big)\big( b_0^-\delta(L_0^-)\otimes\mathbb{I}\otimes\mathbb{I}\big)\big(c_0^-\otimes\mathbb{I}\otimes\mathbb{I}\big)\big(\mathbb{I}\otimes |\mathrm{bpz}^{-1}\rangle\big)\nonumber\\
\lineup = \Big(\big(\langle \mathrm{bpz}|\otimes\mathbb{I}\big)\big(\mathbb{I}\otimes |\mathrm{bpz}^{-1}\rangle\big)\Big)b_0^-\delta(L_0^-)c_0^-\nonumber\\
\lineup = b_0^-\delta(L_0^-)c_0^-.\label{eq:ominvom}
\end{eqnarray}
The final operator acts as the identity on $\Hf$, which establishes that $\omega^{-1}$ is indeed the inverse of $\omega$. Finally we note that the BRST operator $Q$ is BPZ odd:
\begin{eqnarray}
\langle \mathrm{bpz}|(Q\otimes \mathbb{I}+\mathbb{I}\otimes Q) \lineup = 0\\
(Q\otimes \mathbb{I}+\mathbb{I}\otimes Q)|\mathrm{bpz}^{-1}\rangle \lineup = 0.
\end{eqnarray}
This implies that the symplectic form and the Poisson bivector satisfy
\begin{eqnarray}
\langle \omega|(Q\otimes \mathbb{I}+\mathbb{I}\otimes Q) \lineup = 0\label{eq:bosQcyc}\\
(Q\otimes \mathbb{I}+\mathbb{I}\otimes Q)|\omega^{-1}\rangle \lineup = 0.\label{eq:bosQPois}
\end{eqnarray}
Accordingly, we say that $Q$ is {\it cyclic} with respect to the symplectic form $\omega$. Note that because $c_0^-$ does not commute with $Q$, in establishing the first relation it is important to remember that $\omega$ only operates on states in $\Hf$. 

\vspace{.5cm}

\noindent{\it String products $\ell_{g,n}$}: The final and most nontrivial ingredient in the theory are the multi-string products $\ell_{g,n}$. They are Grassmann odd multilinear maps from $n$ copies of $\Hf$ into $\Hf$:
\begin{equation}\ell_{g,n}:\Hf^{\otimes n}\to\Hf.\end{equation}
Thus $\ell_{g,n}$ multiplies $n$ states in $\Hf$ to produce a state in $\Hf$.  The index $g$ refers to the ``genus" of the product. We assume that $\ell_{0,0}$ vanishes, which implies that the theory describes fluctuations of a conformal background at tree level. The 1-string product $\ell_{0,1}$ is identified with the BRST operator $Q$.  $\ell_{g,n}$ carries ghost number $3-2n$. The products satisfy a hierarchy of relations arising from the requirement that the action satisfies the quantum Batalin-Vilkovisky master equation. These relations imply that the products realize a specific algebraic structure, called a {\it quantum $L_\infty$ algebra}.\footnote{Other names include {\it loop homotopy algebra} \cite{Markl} or {\it $IBL_\infty$ algebra} \cite{MuensterSachs}.} A quantum $L_\infty$ algebra is characterized by the following properties:
\begin{description}
\item{(1)} The products are graded symmetric upon interchange of the arguments: 
\begin{equation}\ell_{g,n}(A_1,...,A_i,A_{i+1},...,A_n) =(-1)^{A_iA_{i+1}} \ell_{g,n}(A_1,...,A_{i+1},A_i,...,A_n).\end{equation}
\item{(2)} The products are {\it cyclic} with respect to the symplectic form $\omega$:
\begin{equation}\omega(A_1,\ell_{g,n}(A_2,...A_{n+1})) =-(-1)^{A_1} \omega(\ell_{g,n}(A_1,...A_n),A_{n+1}).\end{equation}
Cyclicity of the BRST operator was already mentioned in \eq{bosQcyc}.
\item{(3)} The products satisfy an infinite hierarchy of Jacobi-like identities called {\it quantum $L_\infty$ relations}. Our main interest is the tadpole amplitude, where the following identities play an important role: 
\begin{eqnarray}
0\lineup =(Q\otimes\mathbb{I}+\mathbb{I}\otimes Q)|\omega^{-1}\rangle \label{eq:qLinf1}\\
0\lineup = Q\ell_{0,2}+\ell_{0,2}(Q\otimes\mathbb{I}+\mathbb{I}\otimes Q)\label{eq:qLinf2}\\
0\lineup = Q\ell_{1,0} +\ell_{0,2}|\omega^{-1}\rangle.\label{eq:qLinf3}
\end{eqnarray}
The first was already given in \eq{bosQPois}.
\end{description}
Since the symplectic form is nondegenerate, the string products may be defined through the vertices
\begin{equation}
\omega(A_1,\ell_{g,n}(A_2,A_3,...,A_{n+1})) \equiv (-1)^{A_1}\{A_1,A_2,...,A_{n+1}\}_g,\label{eq:vertmulti}
\end{equation}
where on the right hand side we have the multilinear string functions as written by Zwiebach \cite{Zwiebach}. The multilinear string functions are graded symmetric upon interchange of any pair string fields, whereas properties (1) and (2) imply that the vertex is graded symmetric only up to a sign from moving string fields through the Grassmann odd products $\ell_{g,n}$. This difference accounts for the sign factor in \eq{vertmulti}. The multilinear string functions are concretely defined by the formula
\begin{equation}
\{A_1,A_2,...,A_n\}_g = \left(\frac{1}{2\pi i}\right)^{d_{g,n}/2}\int_{\mathcal{V}_{g,n}}\langle \Sigma_{g,n}|\frac{b(v)^{d_{g,n}}}{d_{g,n}!}A_1\otimes A_2\otimes...\otimes A_n.\label{eq:bosmulti}
\end{equation}
There are many ingredients here, so let us describe them step by step. The number
\begin{equation}d_{g,n} \equiv 6g-6+2n\end{equation}
is the dimension of the moduli space $\mathcal{M}_{g,n}$ of genus $g$ Riemann surfaces with $n$ punctures. The integration is performed over a subregion $\mathcal{V}_{g,n}$ of the moduli space $\mathcal{M}_{g,n}$. Computing an $n$-point amplitude at genus $g$ requires integrating over the complete moduli space $\mathcal{M}_{g,n}$, and much of this moduli space will be covered by Feynman diagrams composed of lower order vertices connected by propagators. The subregion $\mathcal{V}_{g,n}$ is what is left of the moduli space after these Feynman diagrams have been accounted for. The object $\langle \Sigma_{g,n}|$ is an ``$n$-fold bra ''
\begin{equation}\langle \Sigma_{g,n}|:\Hf^{\otimes n}\to \H^{\otimes 0},\end{equation}
and an example of a {\it surface state}. This means that it is defined by an $n$-point worldsheet correlation function on a genus $g$ Riemann surface with prescribed local coordinates around the punctures. Specifically: 
\begin{equation}
\langle \Sigma_{g,n}(m)|B_1\otimes...\otimes B_n = \Big\langle h_{1}(m)\circ B_1(0)\,...\,h_{n}(m)\circ B_n(0)\Big\rangle^{g}_m.\label{eq:Sigmacorr}
\end{equation}
For clarity we explicitly indicate dependence on the point $m\in\mathcal{V}_{g,n}$. On the right hand side $\big\langle ...\big\rangle^g_m$ denotes a correlation function on a genus $g$ Riemann surface, with operators inside the brackets expressed in a chosen coordinate system on the surface (for example, the uniformization coordinate). The correlation function depends on $m$ through the moduli needed to specify the surface. The states $B_1,...,B_n$ are represented by vertex operators $B_1(0),...,B_n(0)$ which have been inserted at the origin of respective local coordinate disks, denoted by $\xi_1,...,\xi_n$ with $|\xi_i|<1$. The disks are mapped to the surface (in the chosen coordinate system) with the local coordinate maps $h_{1}(m,\xi_1),...,h_{n}(m,\xi_n)$. The local coordinate maps in general depend on $m$. An important condition is that the image of the local coordinate disks on the surface do not overlap. This is because in computing amplitudes we must remove local coordinate disks and glue surfaces together on the resulting holes. Since regions of the surface cannot be ``removed twice," the image of the local coordinate disks cannot overlap. Surface states are BRST invariant:
\begin{equation}\langle\Sigma_{g,n}|(Q\otimes\mathbb{I}^{n-1} + \mathbb{I}\otimes Q\otimes \mathbb{I}^{\otimes n-2} +...+\mathbb{I}^{\otimes n-1}\otimes Q) = 0.
\end{equation}
In terms of the correlation function, the left hand side amounts to surrounding all vertex operators in \eq{Sigmacorr} with a contour integral of the BRST current. This contour integral can then be deformed inside the surface and shrunk to a point, which gives zero. The final ingredient we need for the vertex is the operator $b(v)$, whose role is to turn the surface state $\langle\Sigma_{g,n}|$ into a differential form which can be integrated over $\mathcal{V}_{g,n}$. The operator $b(v)$ is a sum of $b$ ghost insertions acting respectively on each state in the vertex:
\begin{equation}
b(v(m)) \equiv \sum_{i=1}^{n} \mathbb{I}^{\otimes i-1}\otimes b(v^{(i)}(m))\otimes \mathbb{I}^{\otimes n-i},
\end{equation}
where
\begin{equation}
b(v^{(i)}(m)) \equiv \oint_{|\xi_i|=1}\frac{d\xi_i}{2\pi i} v^{(i)}(m,\xi_i) b(\xi_i) +\oint_{|\overline{\xi}_i|=1}\frac{d\overline{\xi}_i}{2\pi i} \overline{v}^{(i)}(m,\overline{\xi}_i) \overline{b}(\overline{\xi}_i), 
\end{equation}
with both contours running counterclockwise in the respective coordinates, and $v^{(i)}(m,\xi_i)$ is a 1-form on $\mathcal{V}_{g,n}$. In coordinates $m^\alpha$ on $\mathcal{V}_{g,n}$, they take the form 
\begin{equation}v^{(i)}(m,\xi_i) = dm^\alpha v^{(i)}_{\alpha}(m,\xi_i).\end{equation}
For simplicity, we assume that the basis 1-forms $dm^\alpha$ commute through the $b$ ghosts and other worldsheet operators without a sign. The coefficients functions $v^{(i)}_\alpha(m,\xi_i)$ are called Schiffer vector fields. We will give more discussion later, but the basic idea is that an infinitesimal change of $\langle \Sigma_{g,n}(m)|$ can be described by removing the image of the local coordinate disks $|\xi_i|<1$ from the Riemann surface, slightly deforming their shape, and then gluing them back. If we change $\langle \Sigma_{g,n}(m)|$ along $m^\alpha$, the corresponding deformation of the disks will be described by vector fields  $v^{(i)}_\alpha(m,\xi_i)$ which are holomorphic in the vicinity of $|\xi_i|=1$. Concretely, a deformation of the $i$th coordinate disk is implemented by a contour integral of the energy momentum tensor 
\begin{equation}
T(v^{(i)}_\alpha(m)) \equiv \oint_{|\xi_i|=1}\frac{d\xi_i}{2\pi i} v^{(i)}_\alpha(m,\xi_i) T(\xi_i) +\oint_{|\overline{\xi}_i|=1}\frac{d\overline{\xi}_i}{2\pi i} \overline{v}^{(i)}_\alpha(m,\overline{\xi}_i) \overline{T}(\overline{\xi}_i) 
\end{equation}
operating on the $i$th state. Defining 
\begin{equation}
T(v_\alpha(m)) \equiv \sum_{i=1}^{n} \mathbb{I}^{\otimes i-1}\otimes T(v^{(i)}_\alpha(m))\otimes \mathbb{I}^{\otimes n-i},
\end{equation}
the Schiffer vector fields are defined so that the following identity holds: 
\begin{equation}\frac{\d}{\d m^\alpha}\langle \Sigma_{g,n}(m)| = -\langle \Sigma_{g,n}(m)|T(v_\alpha(m)).\label{eq:dmT}
\end{equation}
This defines the operator $b(v)$. Given $\langle\Sigma_{g,n}|$ and $b(v)$ we can define a natural $p$-form on $\mathcal{V}_{g,n}$: 
\begin{equation}\langle \Sigma_{g,n}|\frac{b(v)^p}{p!}.\end{equation}
An important property is the BRST identity:
\begin{equation}\langle \Sigma_{g,n}|\frac{b(v)^{p+1}}{(p+1)!}(Q\otimes\mathbb{I}^{n-1}+...+\mathbb{I}^{\otimes n-1}\otimes Q) =(-1)^{p+1} d\left(\langle\Sigma_{g,n}|\frac{b(v)^p}{p!}\right),\label{eq:BRSTid}\end{equation}
where $d$ denotes the exterior derivative on $\mathcal{V}_{g,n}$. See section 7 of \cite{Zwiebach} for the demonstration. This identity plays an important role in the proof of quantum $L_\infty$ relations and the proof that cohomologically trivial states decouple from scattering amplitudes. 

The $n$-string vertex at genus $g$ is completely specified once we have provided the subregion $\mathcal{V}_{g,n}$ of the  moduli space and the local coordinate maps $h_1(m,\xi_1),...h_n(m,\xi_n)$ for each point $m\in \mathcal{V}_{g,n}$. This data is tightly constrained by the requirement that the the products $\ell_{g,n}$ associated to the vertices define a quantum $L_\infty$ algebra. However, there is still some freedom in the definition of each vertex, in particular in the choice of local coordinate maps in the interior of $\mathcal{V}_{g,n}$. Different choices of vertices result in string field theories related by field redefinition \cite{fd}. We will present our construction of the vertices needed for the tadpole amplitude later. See \cite{Zwiebach,Zwiebachmin} for the classical construction based minimal area metrics.

\subsection{Heterotic String Field Theory}
\label{subsec:het}

Heterotic string field theory is the field theory of fluctuations of a background in heterotic string theory. The background is described by a worldsheet conformal field theory: The holomorphic sector is a $c=15$ matter tensored with a $c=-15$ $bc$ and $\beta\gamma$ ghost superconformal field theory, while the antiholomorphic sector is a $c=26$ matter tensored with a $c=-26$ $bc$ ghost conformal field theory. The $\beta\gamma$ ghosts will be bosonized to the $\eta,\xi,e^\phi$ system \cite{FMS}. A string field is an element of the state space $\H$ of this conformal field theory. The state space has three important gradings: ghost number, picture number, and Grassmann parity which distinguishes commuting and anticommuting states. $\H$ contains an important linear subspace, the ``small Hilbert space"
\begin{equation}\H_S\subset\H\end{equation}
composed by states which are independent of the zero mode of the $\xi$ ghost. Therefore states in $\H_S$ satisfy
\begin{equation}\eta A = 0,\ \ \ A\in\H_S,\end{equation}
where $\eta$ is the zero mode of the eta ghost. The full state space $\H$ is called the ``large Hilbert space," since it contains states which depend on the $\xi$ zero mode. $\H$ is a direct sum of state spaces composed of Neveu-Schwarz (NS) and Ramond (R) sector states
\begin{equation}\H = \H_\mathrm{NS}\oplus\H_\mathrm{R}.\end{equation}
We assume that all states in $\H$ are GSO projected. Since the theory includes fermions and spacetime ghosts, we allow states to appear in combinations with Grassmann even or Grassmann odd coefficients. 

Quantum heterotic string field theory is defined in the framework of the Batalin-Vilkovisky formalism. The central object is the quantum master action, whose structure is quite analogous to that of the closed bosonic string:
\begin{eqnarray}
S \lineup = \frac{1}{2!}\Omega(\Phi,Q\Phi) + \frac{1}{3!}\Omega(\Phi,L_{0,2}(\Phi,\Phi)) + \frac{1}{4!}\Omega(\Phi,L_{0,3}(\Phi,\Phi,\Phi)) + \frac{1}{5!}\Omega(\Phi,L_{0,4}(\Phi,\Phi,\Phi,\Phi))+...\nonumber\\
\lineup\ \ \ \ \ \ \ \ \ \ \ \ \ \ \ \ \ \ \ \, +\,  \ \ \ \ \frac{1}{1!} \Omega(\Phi,L_{1,0}) \ \ \ \  \,+\ \ \ \ \frac{1}{2!}\Omega(\Phi,L_{1,1}(\Phi))\ \ \ +\ \ \ \ \frac{1}{3!}\Omega(\Phi,L_{1,2}(\Phi,\Phi))\ \ \, +...\nonumber\\
\lineup\ \ \ \ \ \ \ \ \ \ \ \ \ \ \ \ \ \ \ \ \ \ \ \ \ \ \ \ \ \ \ \ \ \ \ \ \ \ \ \ \ \ \ \ \ \ \ \ \ \ \ \ \ \ \ \ \ \ \ \ \ \ \ \ \ \ \ \ \ \ \ \ \ \ \ \ \ \ \, \  +\ \ \ \ \ \ \ \ \frac{1}{1!}\Omega(\Phi,L_{2,0})\ \ \ \ \ \ \ \, +...\ .\label{eq:masteraction}
\end{eqnarray}
The precise realization of this action depends on the formulation of the Ramond sector. There are two possible approaches. One approach appears in recent work of Kunitomo and Okawa \cite{complete} and subsequent papers \cite{RWaction,KSR,hetR,susyS,Rscatter,susyK}, and is characterized by a Ramond string field at picture $-1/2$ subject to a linear constraint. The constraint is conceptually analogous to the $b_0^-$ and level matching constraints of the closed string, and is well motivated from the supermoduli space perspective. However, a proper treatment requires a more sophisticated approach to the $\beta\gamma$ path integral \cite{revisited}, and we would like to take advantage of standard formulas for $\beta\gamma$ correlators at higher genus \cite{Verlinde} for the purpose of locating spurious poles. Therefore we will follow a different approach devised by Sen \cite{1PIR}, which has a somewhat simpler worldsheet realization.  Assuming Sen's approach, the basic ingredients of the action are as follows:

\vspace{.5cm}

\noindent{\it Dynamical string field $\Phi$}: The state $\Phi$ is the dynamical string field. It is not an arbitrary state in $\H$, but a Grassmann even element of a linear subspace,
\begin{equation}
\Phi\in \Hf\subset\H_S\subset\H,
\end{equation}
characterized by states in the small Hilbert space which satisfy $b_0^-$ and level matching constraints:
\begin{equation}
b_0^- \Phi = L_0^-\Phi = 0.
\end{equation}
There is also a condition on picture number. NS states in $\Hf$ must carry picture $-1$, while Ramond states may carry picture $-1/2$ or $-3/2$. Therefore, the dynamical string field $\Phi$ contains an NS component and a Ramond component:
\begin{equation}
\Phi = \PhiN+\PhiR.
\end{equation}
The NS part carries picture $-1$, while the Ramond part carries both pictures $-1/2$ and $-3/2$.

\vspace{.5cm}

\noindent{\it Symplectic form} $\Omega$. The object $\Omega$ is a symplectic form on $\Hf$. It is graded antisymmetric and nondegenerate. We will often write $\Omega$ as a ``double ket" state 
\begin{equation}
\langle\Omega|:\Hf^{\otimes 2}\to \Hf^{\otimes 0},
\end{equation}
so that 
\begin{equation}
\langle\Omega|A\otimes B = \Omega(A,B).
\end{equation}
The heterotic symplectic form is defined by 
\begin{equation}\langle \Omega| \equiv \langle \omega|(\mathbb{I}-X_0 P_{-3/2})\otimes \mathbb{I}, \Vspace
\end{equation}
where $\omega$ is the symplectic form \eq{bos_om} of the closed bosonic string, with the understanding that the BPZ inner product is computed in the small Hilbert space of the heterotic CFT. Here $P_{-3/2}$ is the projector onto states at picture $-3/2$ and $X_0$ is the zero mode of the picture changing operator:
\begin{equation}X_0 \equiv \oint_{|z|=1}\frac{dz}{2\pi i}\frac{1}{z}X(z),\ \ \ \ X(z) = [Q,\xi(z)].\end{equation}
The PCO is BPZ even:
\begin{equation}\langle\mathrm{bpz}| X_0\otimes\mathbb{I} = \langle \mathrm{bpz}|\mathbb{I}\otimes X_0,\end{equation}
and satisfies
\begin{equation}[L_0^{\pm},X_0] = 0,\ \ \ [b_0^{\pm},X_0]=0,\ \ \ [Q,X_0]= 0.\end{equation}
Therefore $X_0$ preserves the $b_0^-$ and level matching constraints. Generally $X_0$ does not commute through $c_0^-$. However, inside the symplectic form it effectively does, since $\Omega$ always operates on states satisfying $b_0^-$ and level matching constraints. It is important to note that the operator $\mathbb{I}-X_0P_{-3/2}$ does not spoil nondegeneracy of the symplectic form, since it is invertible: 
\begin{equation}
(\mathbb{I}-X_0P_{-3/2})^{-1} = \mathbb{I}+X_0P_{-3/2}.
\end{equation}
In particular, we may construct the Poisson bivector 
\begin{equation}|\Omega^{-1}\rangle \in \H_\Phi^{\otimes 2}\end{equation}
which inverts $\Omega$. To do this we define
\begin{equation}|\omega^{-1}\rangle \equiv \Big[(P_{-1}+P_{-1/2}+P_{-3/2})b_0^-\delta(L_0^-)\otimes\mathbb{I}\Big]|\mathrm{bpz}^{-1}\rangle
\end{equation}
Compared to \eq{bosPois} we must include explicit projections onto the appropriate pictures since otherwise sums over intermediate states result in divergences in loops. The projections are also needed so that $|\omega^{-1}\rangle$ is a state in $\Hf^{\otimes 2}$. The heterotic Poisson bivector is then given by
\begin{equation}
|\Omega^{-1}\rangle = \Big((\mathbb{I}+X_0P_{-3/2})\otimes\mathbb{I}\Big)|\omega^{-1}\rangle. \Vspace
\end{equation}
Following \eq{ominvom}, it is straightforward to show that
\begin{eqnarray}
\big(\langle\Omega|\otimes\mathbb{I}\big)\big(\mathbb{I}\otimes|\Omega^{-1}\rangle\big) \lineup = \big(\mathbb{I}\otimes\langle\Omega|\big)\big(|\Omega^{-1}\rangle\otimes\mathbb{I}\big)\nonumber\\
\lineup = (P_{-1}+P_{-1/2}+P_{-3/2})b_0^-c_0^-\delta(L_0^-).
\end{eqnarray}
The final operator acts as the identity on $\Hf$. We also have
\begin{eqnarray}
\langle \Omega|(Q\otimes \mathbb{I}+\mathbb{I}\otimes Q) \lineup = 0\label{eq:Qcyc}\\
(Q\otimes \mathbb{I}+\mathbb{I}\otimes Q)|\Omega^{-1}\rangle \lineup = 0.\label{eq:QPois}
\end{eqnarray}
Accordingly, we say that $Q$ is {\it cyclic} with respect to the symplectic form $\Omega$. In conclusion, we see that $\Omega$ shares the same algebraic properties as the corresponding symplectic form $\omega$ of the closed bosonic string. By the usual argument, we can vary the free action to obtain the classical linearized equations of motion:
\begin{equation}Q\Phi = 0.\end{equation}
This has a somewhat odd consequence: Because $\Phi$ contains states at picture $-1/2$ and at $-3/2$, the theory includes two copies of the Ramond cohomology. The main point, however, is that the interactions of the string field can be chosen in such a way that one copy of the cohomology decouples from scattering amplitudes. Therefore, the additional states can be effectively ignored. The constrained formulation of \cite{complete} produces the correct cohomology, and may be more suitable for understanding general coordinate invariance \cite{SenSugra} and the relation to supermoduli space, but this will not be a central concern for us.

\vspace{.5cm}

\noindent {\it String Products $L_{g,n}$}. The final ingredient in the theory are the multi-string products $L_{g,n}$. They are Grassmann odd linear maps from $n$ copies of $\Hf$ into $\Hf$:
\begin{equation}L_{g,n}:\Hf^{\otimes n}\to\Hf.\end{equation}
The index $g$ refers to the ``genus" of the product. We assume that $L_{0,0}$ vanishes, and $L_{0,1}$ is identified with the BRST operator $Q$. $L_{g,n}$ carries ghost number $3-2n$. The quantum Batalin-Vilkovisky master equation requires that the products $L_{g,n}$ define a quantum $L_\infty$ algebra. The requisite properties are precisely the same as for the closed bosonic string, with $\ell_{g,n},\omega$ replaced with $L_{g,n},\Omega$.

In principle, the products $L_{g,n}$ could be described by of integration over subspaces of the supermoduli space of super Riemann surfaces \cite{Muenster,OkawaSuperMod}, but for present purposes this is not the most useful characterization. Instead we will describe the products in terms of a configuration of PCOs operating on the products $\ell_{g,n}$ of the closed bosonic string. The PCOs must be chosen in a specific fashion so that $L_{g,n}$ define a quantum $L_\infty$ algebra. A second condition, specific to Sen's formulation of the Ramond sector, is ensuring that the spurious copy of the Ramond cohomology decouples from scattering amplitudes. This is implemented by requiring that all products besides the BRST operator take the form
\begin{equation}L_{g,n} = (\mathbb{I}+X_0 P_{-3/2})C_{g,n},\end{equation}
where the products
\begin{equation}C_{g,n}:\Hf^{\otimes n}\to\Hf\label{eq:C}\end{equation}
have the following properties:
\begin{description}
\item{(a)} $C_{g,n}$ vanishes when multiplying any Ramond state at picture $-3/2$.
\item{(b)} The output of the product $C_{g,n}$ only contains NS states at picture $-1$ and Ramond states at picture $-3/2$.
\item{(c)} $C_{g,n}$ is cyclic with respect to the bosonic symplectic form $\omega$.
\end{description}
With these conditions, one can show that the $n$-string vertex at genus $g$ takes the form
\begin{equation}
\Omega(\Phi,L_{g,n}(\Phi,...,\Phi)) = \omega(\widehat{\Phi},C_{g,n}(\widehat{\Phi},...,\widehat{\Phi})),
\end{equation}
where $\widehat{\Phi}$ is equal to $\Phi$ with the component at picture $-3/2$ set to zero. Therefore, the states at picture $-3/2$ decouple from the action and do not contribute to scattering amplitudes. Note that our presentation differs slightly from Sen's, who formulates the action in terms of $C_{g,n}$ rather than the quantum $L_\infty$ products $L_{g,n}$. For this reason, the  proof of the quantum master equation required slight modification of the standard argument~\cite{SenBV}. Analogous modifications are needed in the proof of gauge invariance in open string field theory \cite{RWaction,KSR}.

\section{One Loop Tadpole in Closed Bosonic String Field Theory}
\label{sec:bosonictad}

To compute heterotic tadpole we must first specify all data associated to integration over the bosonic moduli space of the 1-punctured torus. This effectively requires computing the 1-loop tadpole in closed bosonic string field theory.

To compute this amplitude we must fix a gauge. We choose Siegel gauge:
\begin{equation}b_0^+\Phi = 0.\end{equation}
The amplitude is then given by\footnote{We do not attempt to fix the normalization of the amplitude, but the sign is natural (particularly at higher points) to ensure that the amplitude is graded symmetric. For on-shell amplitudes the asymptotic states are Grassmann even, and the sign drops~out.} 
\begin{equation}
\mathcal{A}_{1,1}^{\mathrm{bos}}(\Phi_1) = -(-1)^{\Phi_1}\omega\left(\Phi_1,\ell_{0,2}\left(\frac{b_0^+}{L_0^+}\otimes \mathbb{I}\right)|\omega^{-1}\rangle\right)+ (-1)^{\Phi_1}\omega\left(\Phi_1, \ell_{1,0}\right),\label{eq:hettad}
\end{equation}
where $\Phi_1$ is an (off-shell) state in $\Hf$. The first term arises from the Feynman diagram where two legs of the tree-level cubic vertex are tied together with a Siegel gauge propagator. The second term comes from the fundamental 1-loop tadpole vertex, which can be seen as a counterterm which ``renormalizes" the tadpole amplitude computed from the tree-level action. To see that the expression is correct, let us check that BRST exact states decouple. The amplitude is characterized by contracting $\Phi_1$ with the 0-string product 
\begin{equation}
A_{1,0} =  - \ell_{0,2}\left(\frac{b_0^+}{L_0^+}\otimes \mathbb{I}\right)|\omega^{-1}\rangle+\ell_{1,0}. \label{eq:A10}
\end{equation}
Decoupling of BRST exact states is equivalent to the statement that $A_{1,0}$ is BRST invariant. Let us check this: 
\begin{equation}
Q A_{1,0} = Q \ell_{1,0} - Q\ell_{0,2}\left(\frac{b_0^+}{L_0^+}\otimes \mathbb{I}\right)|\omega^{-1}\rangle.
\end{equation}
Recalling the quantum $L_\infty$ relations \eq{qLinf1} and \eq{qLinf2} this simplifies to
\begin{eqnarray}
Q A_{1,0} \lineup = Q \ell_{1,0} + \ell_{0,2}(Q\otimes\mathbb{I}+\mathbb{I}\otimes Q)\left(\frac{b_0^+}{L_0^+}\otimes \mathbb{I}\right)|\omega^{-1}\rangle\nonumber\\
\lineup = Q \ell_{1,0} + \ell_{0,2}\left(Q\frac{b_0^+}{L_0^+}\otimes \mathbb{I}-\left(\frac{b_0^+}{L_0^+}\otimes\mathbb{I}\right)(\mathbb{I}\otimes Q)\right)|\omega^{-1}\rangle\nonumber\\
\lineup = Q \ell_{1,0} + \ell_{0,2}\left(Q\frac{b_0^+}{L_0^+}\otimes \mathbb{I}+\left(\frac{b_0^+}{L_0^+}\otimes\mathbb{I}\right)(Q\otimes \mathbb{I})\right)|\omega^{-1}\rangle\nonumber\\
\lineup = Q \ell_{1,0} + \ell_{0,2}\left(\left[Q,\frac{b_0^+}{L_0^+}\right]\otimes \mathbb{I}\right)|\omega^{-1}\rangle.
\end{eqnarray}
The commutator of $Q$ with $b_0^+/L_0^+$ is the identity operator up to contributions from the boundary of moduli space which we ignore. Therefore
\begin{equation}
Q A_{1,0} = Q \ell_{1,0} + \ell_{0,2}|\omega^{-1}\rangle,
\end{equation}
which vanishes according to the quantum $L_\infty$ relation \eq{qLinf3}.

\begin{figure}
\begin{center}
\resizebox{2.2in}{2.5in}{\includegraphics{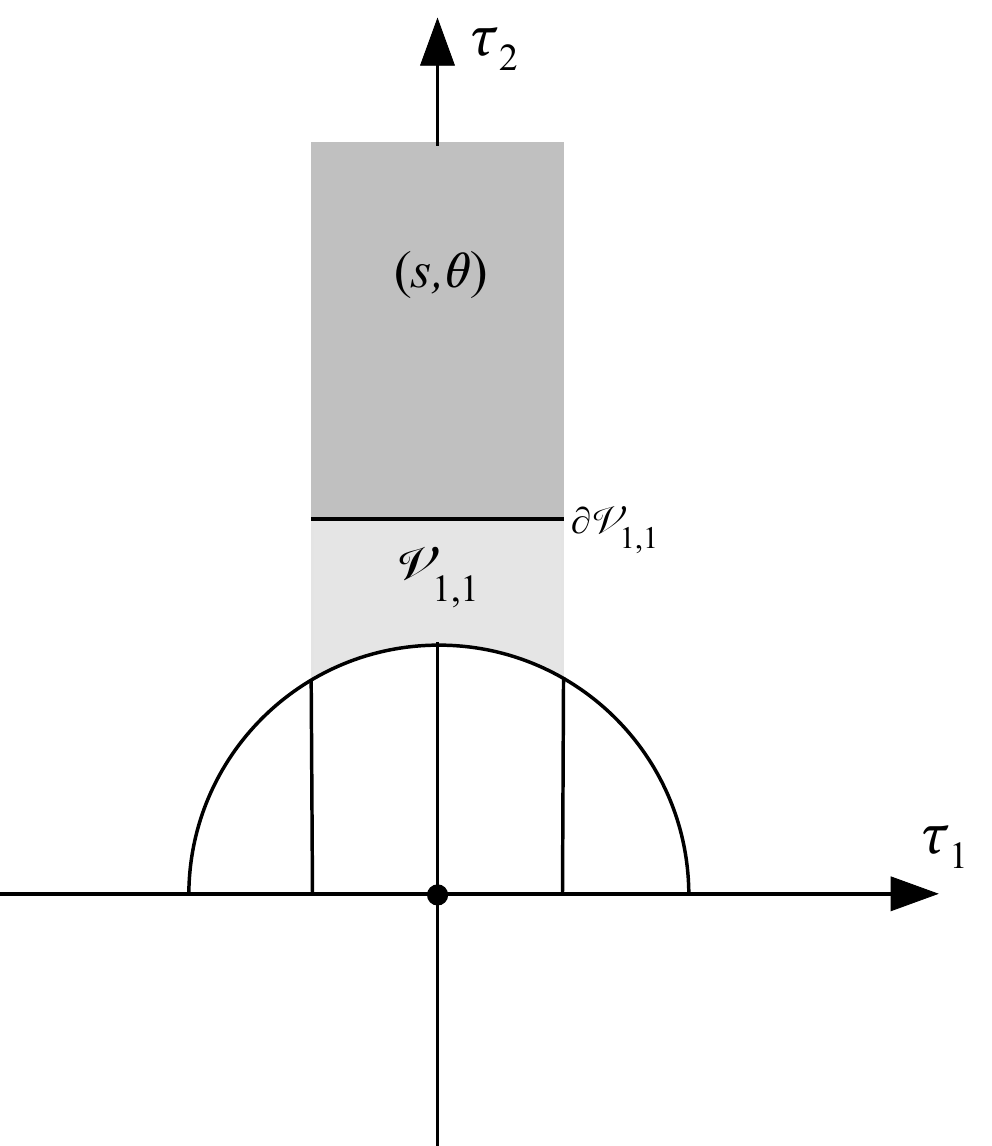}}
\end{center}
\caption{\label{fig:tadmod} The moduli space of a 1-punctured torus, as represented by the fundamental region of the standard modular parameter $\tau = \tau_1+i\tau_2$. The subregion shaded in dark grey is associated with the tree level cubic vertex with two legs attached to a propagator tube of length $s$ and twist angle $\theta$. The subregion shaded light grey is associated with the fundamental 1-loop tadpole vertex. The curve $\d \mathcal{V}_{1,1}$ representing the interface between these two regions depends in detail on the definition of the tree level cubic vertex, and part of our task will be to determine this curve.}
\end{figure}

It is helpful to understand this derivation at the level of the surface states defining the products. The amplitude can be written
\begin{equation}
\mathcal{A}_{1,1}^\mathrm{bos}(\Phi_1)
 = -\int_0^\infty ds\int_{-\pi}^\pi \frac{d\theta}{2\pi} \langle \Sigma_{0,3}|\left(\mathbb{I}\otimes b_0^+b_0^-e^{-sL_0^++i\theta L_0^-}\otimes\mathbb{I}\right)\Phi_1\otimes |\mathrm{bpz}^{-1}\rangle +\frac{1}{2\pi i}\int_{\mathcal{V}_{1,1}}\langle\Sigma_{1,1}|\frac{b(v)^2}{2!} \Phi_1.
\end{equation}
We expanded the propagator and level matching projector into an integral over a Schwinger parameter $s\in[0,\infty]$ and a twist angle $\theta\in[-\pi,\pi]$. The operator $e^{-sL_0^++i\theta L_0^-}$ can be visualized as a cylindrical tube of worldsheet of length $s$ twisted by an angle $\theta$. The moduli space of the 1-punctured torus is two dimensional, and may be represented by the fundamental region of the modular parameter $\tau$ of the torus, as shown in figure \ref{fig:tadmod}. The integration variables $s,\theta$ are coordinates on the part of the moduli space of the 1-punctured torus produced by the tree level cubic vertex with two legs connected by a propagator. Since the limit $s\to\infty$ represents a degenerate torus whose handle acquires infinite length, we can anticipate that the coordinates $s,\theta$ will cover the part of the moduli space shown in figure \ref{fig:tadmod}. The remainder of the moduli space defines $\mathcal{V}_{1,1}$, and is associated with the fundamental 1-loop tadpole vertex. Let us see how BRST exact state decouple in this language. Suppose $\Phi_1=Q\Lambda$ and consider the propagator term. Using BRST invariance of the surface state $\langle \Sigma_{0,3}|$ and $|\mathrm{bpz}^{-1}\rangle$ we obtain
\begin{eqnarray}
\lineup\!\!\!\!\!\!\!\!\!\! -\int_0^\infty ds\int_{-\pi}^\pi \frac{d\theta}{2\pi} \langle \Sigma_{0,3}|\left(\mathbb{I}\otimes b_0^+b_0^-e^{-sL_0^++i\theta L_0^-}\otimes\mathbb{I}\right)Q\Lambda\otimes |\mathrm{bpz}^{-1}\rangle\nonumber\\
\lineup\ \ \ \ \ \   = \int_0^\infty ds\int_{-\pi}^\pi \frac{d\theta}{2\pi} \langle \Sigma_{0,3}|\left(\mathbb{I}\otimes[Q, b_0^+b_0^-e^{-sL_0^++i\theta L_0^-}]\otimes\mathbb{I}\right)\Lambda\otimes |\mathrm{bpz}^{-1}\rangle\nonumber\\
\lineup\ \ \ \ \ \  = \int_0^\infty ds\int_{-\pi}^\pi \frac{d\theta}{2\pi} \langle \Sigma_{0,3}|\left(\mathbb{I}\otimes (L_0^+b_0^--b_0^+L_0^-)e^{-sL_0^++i\theta L_0^-}\otimes\mathbb{I}\right)\Lambda\otimes |\mathrm{bpz}^{-1}\rangle\nonumber\\
\lineup\ \ \ \ \ \ = \int_0^\infty ds\int_{-\pi}^\pi \frac{d\theta}{2\pi} \left[-\frac{\d}{\d s}\langle \Sigma_{0,3}|\left(\mathbb{I}\otimes b_0^-e^{-sL_0^++i\theta L_0^-}\otimes\mathbb{I}\right)-i\frac{\d}{\d\theta}\langle\Sigma_{0,3}|\left(\mathbb{I}\otimes b_0^+e^{-sL_0^++i\theta L_0^-}\otimes\mathbb{I}\right)\right]\Lambda\otimes |\mathrm{bpz}^{-1}\rangle.\nonumber\\
\end{eqnarray}
As explained in \eq{LmdLm}, the integral of the $\theta$ total derivative vanishes because $L_0^-$ takes only integer eigenvalues. We will also see that $\theta=\pm\pi$ correspond to boundaries of the fundamental region of the moduli space which are identified. Therefore only the the $s$ total derivative contributes, giving
\begin{eqnarray}
\lineup -\int_0^\infty ds\int_{-\pi}^\pi \frac{d\theta}{2\pi} \langle \Sigma_{0,3}|\left(\mathbb{I}\otimes b_0^+b_0^-e^{-sL_0^++i\theta L_0^-}\otimes\mathbb{I}\right)Q\Lambda\otimes |\mathrm{bpz}^{-1}\rangle\nonumber\\
\lineup\ \ \ \ \ \ \ \ \ \ \ \ \ \ = \int_{-\pi}^\pi \frac{d\theta}{2\pi} \langle \Sigma_{0,3}|\left(\mathbb{I}\otimes b_0^-e^{i\theta L_0^-}\otimes\mathbb{I}\right)\Lambda\otimes |\mathrm{bpz}^{-1}\rangle.
\end{eqnarray}
Now focus on the fundamental tadpole vertex. Using \eq{BRSTid} and Stokes' theorem we obtain
\begin{eqnarray}
\frac{1}{2\pi i}\int_{\mathcal{V}_{1,1}}\langle\Sigma_{1,1}|\frac{b(v)^2}{2!} Q\Lambda\lineup = \frac{1}{2\pi i}\int_{\mathcal{V}_{1,1}}d\Big(\langle\Sigma_{1,1}|b(v) \Lambda\Big)\nonumber\\
\lineup = \frac{1}{2\pi i}\int_{\d\mathcal{V}_{1,1}}\langle\Sigma_{1,1}|b(v) \Lambda .\end{eqnarray}
We require that $\d\mathcal{V}_{1,1}$ corresponds the $s=0$ boundary of the propagator region of the moduli space. On this boundary we can use the twist angle $\theta$ as a natural coordinate, and write
\begin{equation}
\frac{1}{2\pi i}\int_{\mathcal{V}_{1,1}}\langle\Sigma_{1,1}|\frac{b(v)^2}{2!} Q\Lambda = \frac{1}{2\pi i}\int_{-\pi}^\pi d\theta\, \langle \Sigma_{1,1}(\theta)|b(v_\theta(\theta)) \Lambda,
\end{equation} 
where $v_\theta(\theta)$ is the Schiffer vector field satisfying
\begin{equation}
\frac{d}{d\theta}\langle \Sigma_{1,1}(\theta)| = -\langle \Sigma_{1,1}(\theta)| T(v_\theta(\theta)).
\end{equation}
The tadpole amplitude for a BRST exact state therefore takes the form 
\begin{equation}
\mathcal{A}_{1,1}^\mathrm{bos}(Q\Lambda) = \frac{1}{2\pi i}\left[-\int_{-\pi}^\pi d\theta\, \langle \Sigma_{0,3}|\left(\mathbb{I}\otimes (-ib_0^-)e^{i\theta L_0^-}\otimes\mathbb{I}\right)\Lambda\otimes |\mathrm{bpz}^{-1}\rangle+\int_{-\pi}^\pi d\theta\, \langle \Sigma_{1,1}(\theta)|b(v_\theta(\theta)) \Lambda\right].\label{eq:Qcan1}
\end{equation}
These terms must cancel. To see how this cancellation can occur, note that the $b$ ghost insertion in the first term can be rewritten
\begin{equation}
\langle \Sigma_{0,3}|\left(\mathbb{I}\otimes (-ib_0^-)e^{i\theta L_0^-}\otimes\mathbb{I}\right)\Lambda\otimes |\mathrm{bpz}^{-1}\rangle = \langle \Sigma_{0,3}|\left(\mathbb{I}\otimes e^{i\theta L_0^-}\otimes\mathbb{I}\right)b(v_\theta(\theta))\Lambda\otimes |\mathrm{bpz}^{-1}\rangle,\label{eq:bSchifbm}
\end{equation}
where $v_\theta(\theta)$ is the Schiffer vector field corresponding to $d/d\theta$ for this surface. To see why this is the case, note that from the perspective of surface state conservation laws, the $b$ ghost is equivalent to the energy momentum tensor. If we replace $b$ with the energy momentum tensor above, the two sides of the equation are equal to $-d/d\theta$ of the surface state, and in particular are equal to each other. Therefore \eq{Qcan1} will be zero if we assume:
\begin{equation}
\langle\Sigma_{0,3}|\Big(\mathbb{I}\otimes e^{i\theta L_0^-}\otimes\mathbb{I}\Big)\Big(\mathbb{I}\otimes|\mathrm{bpz}^{-1}\rangle\Big) = \langle\Sigma_{1,1}(\theta)|.
\end{equation}
This fixes the form of the 1-loop tadpole vertex at the boundary of $\mathcal{V}_{1,1}$. To construct the amplitude, we therefore proceed as follows: First we must define a suitable cubic vertex. Second, we characterize the genus 1 surface states obtained by gluing two local coordinate disks of the cubic vertex with a propagator; the gluing is achieved through the standard plumbing fixture relation. This determines the missing region of the moduli space $\mathcal{V}_{1,1}$ and the local coordinate of the 1-loop tadpole surface state on $\d\mathcal{V}_{1,1}$. We may then make a continuous choice of local coordinate on the remainder of $\mathcal{V}_{1,1}$ and construct the appropriate $2$-form to be integrated over $\mathcal{V}_{1,1}$. This specifies all data needed for evaluation of the off-shell 1-loop tadpole amplitude of closed bosonic string field theory.

\subsection{Elementary Cubic Vertex}

We begin by defining the elementary cubic vertex:
\begin{eqnarray}
\omega(A_1,\ell_{0,2}(A_2,A_3)) \lineup = (-1)^{A_1}\langle \Sigma_{0,3}|A_1\otimes A_2\otimes A_3 \nonumber\\
\lineup =(-1)^{A_1} \Big\langle\, f_0\circ A_1(0)\, f_1\circ A_2(0)\, f_\infty\circ A_3(0)\,\Big\rangle,
\end{eqnarray}
where in the final expression we have a correlation function on the complex plane. The global coordinate on the complex plane will be denoted $z$. The vertex operators $A_1(0),A_2(0)$ and $A_3(0)$ are inserted at the origin of respective local coordinate disks $\xi_0,\xi_1,\xi_\infty$ with $|\xi_i|<1$, and the local coordinate maps
\begin{equation}f_0(\xi_0),\ \ \ f_1(\xi_1),\ \ \ f_\infty(\xi_\infty),\end{equation}
transform the disks into the global coordinate $z$. We take the vertex operators to be inserted at $0$, $1$ and $\infty$, which means that the local coordinate maps satisfy
\begin{equation}
f_0(0) = 0,\ \ \ f_1(0) = 1,\ \ \ f_\infty(0) = \infty.\label{eq:cubpunct}
\end{equation}
To simplify the computation of the tadpole amplitude we will assume that $f_0,f_1,f_\infty$ take the form of $SL(2,\mathbb{C})$ transformations. In particular we will not use a Witten-type vertex, so off-shell amplitudes will be different from the canonical ones defined by minimal area metrics. In principle we are free to use any $SL(2,\mathbb{C})$ maps in defining the cubic vertex, but we would need to perform a sum over permutations exchanging vertex operators between punctures to ensure that the vertex is symmetric. However, it is simpler to choose local coordinate maps that result in a symmetric vertex without requiring a sum of permutations. Assuming $SL(2,\mathbb{C})$ local coordinate maps, the resulting cubic vertex is unique up to an overall scaling of the local coordinate disks, as we now describe.

To describe the symmetry of the cubic vertex, we introduce an $S_3$ subgroup of $SL(2,\mathbb{C})$ which interchanges the punctures at $0,1$ and $\infty$. The subgroup has generators
\begin{equation}a(z) = 1-z,\ \ \ b(z) = 1/z.\end{equation}
The group element $a$ interchanges the punctures at $0$ and $1$, leaving the puncture at $\infty$ fixed, while $b$ interchanges the punctures at $0$ and $\infty$ leaving the puncture at $1$ fixed. To verify that $a$ and $b$ generate an $S_3$ subgroup of $SL(2,\mathbb{C})$, it is sufficient to check that $a$ and $b$ satisfy the identities of the presentation of $S_3$:
\begin{eqnarray}
a\circ a(z) \lineup = e(z)\\ 
b\circ b(z) \lineup = e(z)\\
a\circ b\circ a\circ b\circ a\circ b(z) \lineup = e(z),
\end{eqnarray}
where $e(z)=z$ is the identity map. The group element that interchanges the punctures at $1$ and $\infty$, leaving $0$ fixed, takes the form
\begin{equation}
a\circ b\circ a(z) = \frac{z}{z-1}.
\end{equation}
Since $SL(2,\mathbb{C})$ preserves correlation functions on the plane, we have
\begin{equation}
\Big\langle\, f_0\circ A_1(0)\, f_1\circ A_2(0)\, f_\infty\circ A_3(0)\,\Big\rangle = \Big\langle\,\sigma\circ f_0\circ A_1(0)\, \sigma\circ f_1\circ A_2(0)\, \sigma\circ f_\infty\circ A_3(0)\,\Big\rangle,
\end{equation}
where $\sigma(z)\in S_3$. The map $\sigma$ will permute the locations of the punctures, but generally it will not permute the local coordinate maps. That is, $\sigma\circ f_i$ will in general be different from $f_{\sigma(i)}$. However, if they happen to be equivalent, we would have 
\begin{equation}
\Big\langle\, f_0\circ A_1(0)\, f_1\circ A_2(0)\, f_\infty\circ A_3(0)\,\Big\rangle = \Big\langle\, f_{\sigma(0)}\circ A_1(0)\, f_{\sigma(1)}\circ A_2(0)\, f_{\sigma(\infty)}\circ A_3(0)\,\Big\rangle,
\end{equation}
and the vertex would be symmetric. Note that
\begin{equation}e^{i\theta L_0^-}|A\rangle = e^{i\theta}\circ A(0)|0\rangle,\end{equation}
where $e^{i\theta}\circ$ denotes the conformal transformation $z\to e^{i\theta}z$. Therefore vertex operators of level matched states are invariant under conformal transformations which rotate by a phase. Since the cubic vertex is always evaluated on level matched states, we only need to require that $\sigma\circ f_i$ and $f_{\sigma(i)}$ are equal up to a phase rotation of the local coordinate disk:
\begin{equation}\sigma \circ f_i(\xi) = f_{\sigma(i)}(e^{i\theta_{i,\sigma}} \xi)\label{eq:fisym}\end{equation}
This identity characterizes the local coordinate maps of a symmetric cubic vertex.

\begin{figure}
\begin{center}
\resizebox{2.3in}{2.5in}{\includegraphics{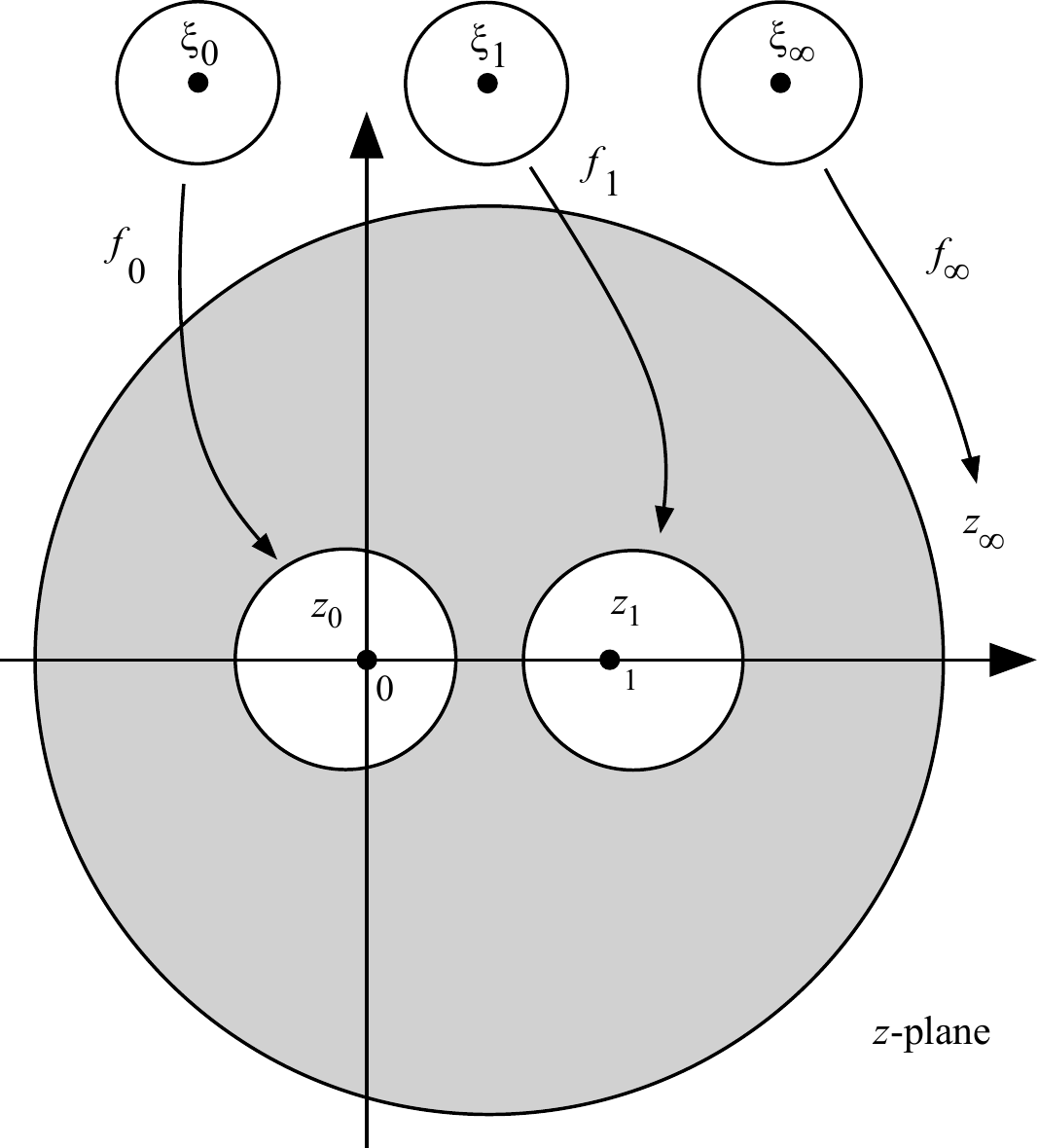}}
\end{center}
\caption{\label{fig:cubicvert} Picture of the surface state defining the cubic vertex. The three circles above represent the local coordinate disks $\xi_0,\xi_1,\xi_\infty$ for the three punctures. These are transformed into regions $z_0,z_1,z_\infty$ in the complex plane by the respective local coordinate maps $f_0,f_1,f_\infty$. The shaded region is the ``interior" of the vertex, that is, the part of the surface which is not inside any local coordinate. Since the maps $f_0,f_1,f_\infty$ are $SL(2,\mathbb{C})$, the image of the local coordinate disks are circles. The picture shows the vertex for a generic value of the stub parameter $\varpi>0$. Towards $\varpi=0$, the exterior circle shrinks and the interior circles grow until they touch on the real axis.}
\end{figure}

To solve this identity let us first consider $f_0$ and the permutation which switches the punctures at $1$ and $\infty$. This permutation should have no effect on the vertex operator inserted with $f_0$, so we should have
\begin{equation}
a\circ b\circ a\circ f_0(\xi) = \frac{f_0(\xi)}{f_0(\xi)-1} = f_0(e^{i\theta_{0,aba}}\xi). \label{eq:f0sym}
\end{equation}
The general $SL(2,\mathbb{C})$ map that preserves the origin takes the form 
\begin{equation}
f_0(\xi) = \frac{\xi}{\alpha \xi+\beta}.
\end{equation}
Plugging this into \eq{f0sym} implies
\begin{equation}
f_0(\xi) = \frac{2\xi}{\xi+2\beta},
\end{equation}
where $\beta$ is an undetermined parameter and the required phase is $e^{i\theta_{0,aba}}=-1$. Without loss of generality we can assume that $\beta$ is real and positive, since a phase can be absorbed into a trivial phase rotation of a level matched state. Next we define $f_1$ and $f_\infty$ by the appropriate permutations of $f_0$:
\begin{eqnarray}
f_1(\xi)\lineup \equiv a\circ f_0(\xi) = \frac{2\beta -\xi}{2\beta+\xi}\\
f_\infty(\xi)\lineup \equiv b\circ f_0(\xi) = \frac{\xi+2\beta}{2\xi}.
\end{eqnarray}
One can then verify that the generators of $S_3$ act on the local coordinate maps in a manner consistent with \eq{fisym}:
\begin{eqnarray}
a\circ f_0(\xi) \lineup = f_1(\xi),\ \ \ \ \ \ \ \ \,b\circ f_0(\xi) = f_\infty(\xi)\\
a\circ f_1(\xi) \lineup = f_0(\xi),\ \ \ \ \ \ \ \ \,b\circ f_1(\xi) = f_1(-\xi)\\
a\circ f_\infty(\xi)\lineup = f_\infty(-\xi),\ \ \ \ b\circ f_\infty(\xi) = f_0(\xi).
\end{eqnarray}
Since all elements of $S_3$ can be obtained by composing the generators, this establishes \eq{fisym} and symmetry of the cubic vertex.

An important condition is that the image of the local coordinate disks do not overlap in the global coordinate~$z$. One may verify that this is the case if the constant $\beta$ satisfies the inequality
\begin{equation}3\leq 2\beta.\end{equation}
It is convenient to solve this inequality by writing $\beta$ in the form
\begin{equation}
\beta = \frac{3}{2}e^{\varpi},
\end{equation}
where $\varpi$ is real and positive number called the {\it stub parameter}. Therefore we may express the local coordinate maps
\begin{eqnarray}
z_0=f_0(\xi_0)\lineup = \frac{2e^{-\varpi}\xi_0}{e^{-\varpi}\xi_0+3}\\
z_1=f_1(\xi_1)\lineup = \frac{3-e^{-\varpi}\xi_1}{3+e^{-\varpi}\xi_1}\label{eq:f1}\\
z_\infty= f_\infty(\xi_\infty)\lineup = \frac{e^{-\varpi}\xi_\infty+3}{2e^{-\varpi}\xi_\infty},
\end{eqnarray}
where $z_0,z_1,z_\infty$ are the image of the local coordinate disks $\xi_0,\xi_1,\xi_\infty$ in the $z$-plane. See figure \ref{fig:cubicvert}. Note that the effect of the stub parameter is to simply rescale the local coordinate disks before conformal transformation to the complex plane. Since
\begin{equation}
e^{-\varpi L_0^+}|A\rangle = e^{-\varpi}\circ A(0)|0\rangle,
\end{equation}
the cubic vertex with $\varpi>0$ is related to the cubic vertex with $\varpi=0$ through
\begin{equation}
\langle \Sigma_{0,3} (\varpi)| = \langle \Sigma_{0,3}(\varpi=0)|e^{-\varpi L_0^+}\otimes e^{-\varpi L_0^+}\otimes e^{-\varpi L_0^+}.
\end{equation}
The operator $e^{-\varpi L_0^+}$ can be visualized  as a tube of worldsheet of length $\varpi$, called a ``stub," which here is attached to every leg of the cubic vertex. Stubs also appear in the cubic vertex defined by the minimal area prescription, where canonically $\varpi=\pi$.

\subsection{Propagator Diagram}

\begin{figure}
\begin{center}
\resizebox{2.5in}{2.4in}{\includegraphics{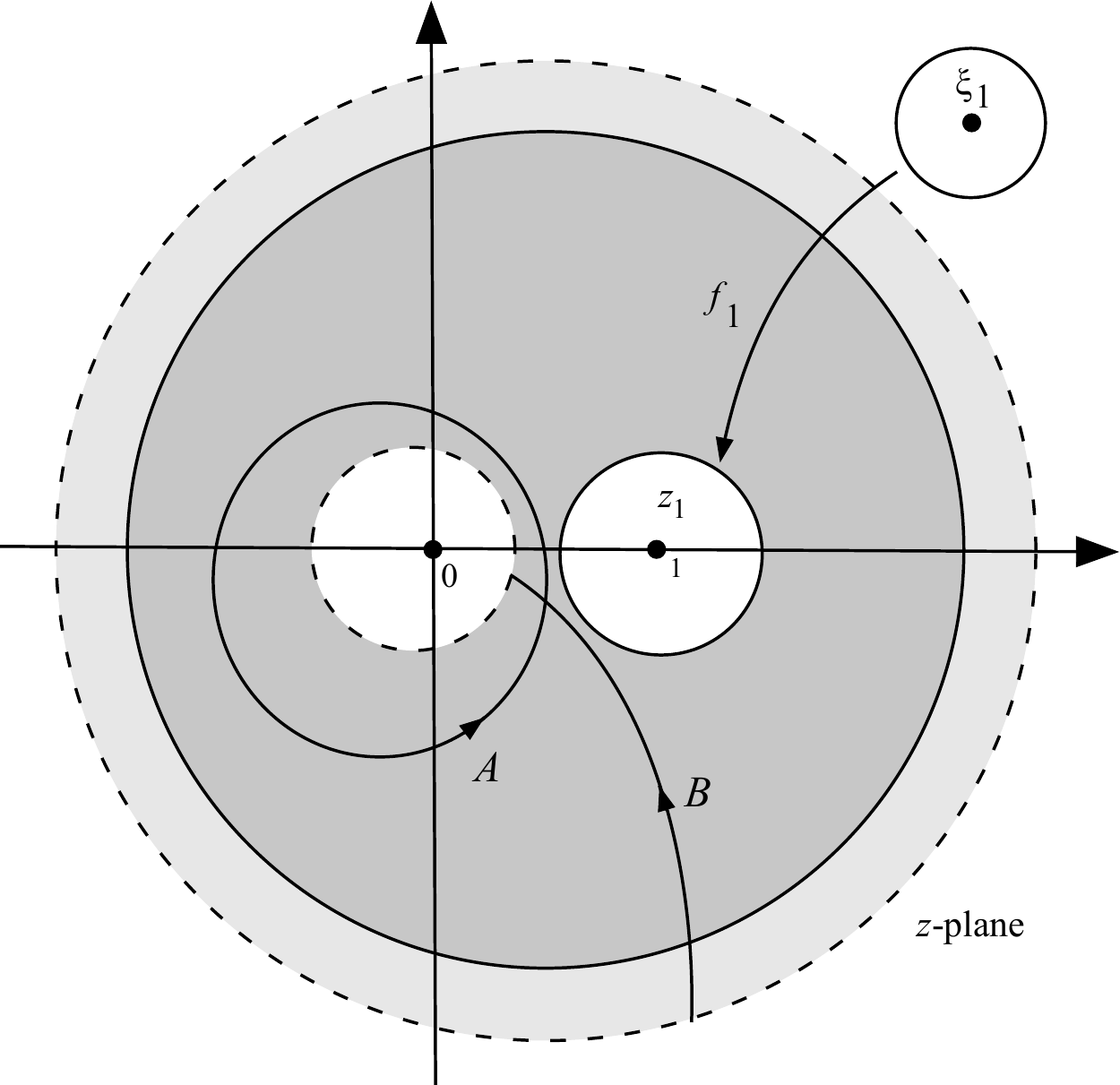}}
\end{center}
\caption{\label{fig:torusz} Picture of the surface state \eq{propsurface} in the $z$-plane. The local coordinate $\xi_1$, representing the external state of the tadpole, is mapped to the $z$-plane just as in the cubic vertex. The inner and outer dashed circles are glued together (with a twist) to create a torus. The dark grey region represents the part of the torus created by the cubic vertex, and the lighter grey region is the part of the torus created by the propagator tube. A choice of $A$ and $B$ cycles on the torus is also shown.}
\end{figure}

The next step is to compute the contribution to the tadpole from the cubic vertex and propagator. This diagram is characterized by the surface state 
\begin{equation}
\langle \Sigma_{0,3}|\Big(\mathbb{I}\otimes e^{-sL_0^+ +i\theta L_0^-}\otimes\mathbb{I}\Big)\Big(\mathbb{I}\otimes |\mathrm{bpz}^{-1}\rangle\Big).\label{eq:propsurface}
\end{equation}
The net effect of the second two factors is to glue a pair local coordinate disks in the cubic vertex with a plumbing fixture. If the glued coordinates are taken to be $\xi_0$ and $\xi_\infty$, the pluming fixture relation reads 
\begin{equation}\xi_0\xi_\infty = e^{-s+i\theta}.\end{equation}
In the $z$ coordinate on the complex plane, the plumbing fixture relation implies an identification between points in the region $z_0$ and $z_\infty$:
\begin{equation}
z_\infty = \frac{1+A(s,\theta) z_0}{2-z_0},\label{eq:zplumb}
\end{equation}
where
\begin{equation}
A(s,\theta) = \frac{9 e^{2\varpi +s -i\theta}-1}{2}.\label{eq:A}
\end{equation}
The result is a genus 1 surface state with 1 puncture. A picture of this surface state in the $z$ plane is shown in figure \ref{fig:torusz}, along with a choice of $A$ and $B$ homology cycles.

The $z$ coordinate, however, is a fairly unnatural parameterization of the torus. We would like to describe the surface state \eq{propsurface} in the uniformization coordinate, defined by the complex plane $w$ with the identifications
\begin{equation}w\sim w+1,\ \ \ w\sim w+\tau,\label{eq:wid}\end{equation}
where $\tau$ is the standard modular parameter of the torus. It is also important to understand how the Schwinger length $s$ and twist angle $\theta$ are related to the modular parameter $\tau$. In this way, we will be able to see what part of the moduli space is covered by the propagator and cubic vertex and what part will need to be covered by the fundamental tadpole vertex. To address these questions, we must construct the holomorphic 1-form on the torus expressed in the $z$ coordinate:
\begin{equation}\omega(z)dz.\end{equation}
Assuming the standard normalization
\begin{equation}
\int_A \omega(z)dz = 1,\label{eq:omA}
\end{equation}
the holomorphic 1-form is unique. We may then identify $\tau$ using the formula 
\begin{equation}\tau = \int_B dz\,\omega(z).\end{equation}
The transformation to the uniformization coordinate is given by the Abel map
\begin{equation}
w=W(z) = \int_1^z dz\,\omega(z).\label{eq:Abel}
\end{equation}
The lower limit of the integral has been chosen so that the puncture will be located at $w=0$ in the uniformization coordinate. Since $\omega(z)$ is holomorphic, the integral will be independent of integration contour in the $z$ plane modulo the identifications \eq{wid}.

The holomorphic 1-form should be single valued on the torus. Therefore it should satisfy
\begin{equation}\omega(z_0)dz_0 = \omega(z_\infty)dz_\infty,\end{equation}
where $z_0$ and $z_\infty$ are related through the plumbing fixture as given in \eq{zplumb}. This amounts to a functional equation for $\omega(z)$:
\begin{equation}\omega(z) = \frac{1+2A}{(2-z)^2}\,\omega\!\left(\frac{1+Az}{2-z}\right).\label{eq:funceq}
\end{equation}
Through some guesswork, we were lead to a solution in the form of the logarithmic derivative of an $SL(2,\mathbb{C})$ transformation:
\begin{equation}
\omega(z)dz = \frac{1}{2\pi i}d\left(\ln\frac{z-R_-}{z-R_+}\right),
\end{equation}
where
\begin{eqnarray}
R_- \lineup = -\frac{1}{2}\left(A-2 - A \sqrt{1-\frac{4}{A}}\right)\\
R_+\lineup = -\frac{1}{2}\left(A-2 + A \sqrt{1-\frac{4}{A}}\right).
\end{eqnarray}
The holomorphic 1-form has poles in the $z$ coordinate for $z=R_\pm$. One may check that the pole at $R_-$ always appears inside the region $z_0$, and the pole at $R_+$ always appears inside $z_\infty$ with the propagator region excluded. Therefore, poles never appear in the region in the $z$-plane that define the torus itself. Computing the exterior derivative we find
\begin{equation}
\omega(z)dz = \frac{1}{2\pi i} \frac{R_--R_+}{(z-R_-)(z-R_+)} dz.
\end{equation}
Computing the contour integral along the $A$ cycle we pick up a residue from the pole at $z=R_-$:
\begin{equation}
\int_A dz\, \omega(z) = 2\pi i\left.\left(\frac{1}{2\pi i}\frac{R_--R_+}{z-R_+}\right)\right|_{z=R_-} = 1.
\end{equation}
So the holomorphic 1-form is correctly normalized.

It is interesting to see why the holomorphic 1-form can be derived in this fashion. Suppose that $\omega(z)dz$ is expressed as the exterior derivative of some $F(z)$. The functional equation \eq{funceq} can then be rewritten
\begin{equation}F(z) = F(P(z))+c,\label{eq:Ffunc}\end{equation}
where $P(z)$ is the $SL(2,\mathbb{C})$ transformation \eq{zplumb} defining the plumbing fixture relation, and $c$ is a constant of integration. Let us further reexpress $F(z)$ in the form
\begin{equation}F(z) = \ln S(z), \end{equation}
for some function $S(z)$. Then the functional equation for $S(z)$ takes the form. 
\begin{equation}C S(z) = S(P(z)),\end{equation} 
where $C$ is related to $c$ in the first step. This equation determines an $SL(2,\mathbb{C})$ transformation $S(z)$ provided that  that $P(z)$ is in the same conjugacy class as dilatation by some constant $C$. One can readily verify that this is the case. The value of $C$ is determined since the $SL(2,\mathbb{C})$ matrices of $C$ and $P$ must share a common trace. Note that $P(z)$ is not in the same conjugacy class as a translation. This is why we needed the logarithm; equation \eq{Ffunc} does not have a solution for $F(z)$ as an $SL(2,\mathbb{C})$ transformation. 

\begin{figure}
\begin{center}
\resizebox{6in}{1.9in}{\includegraphics{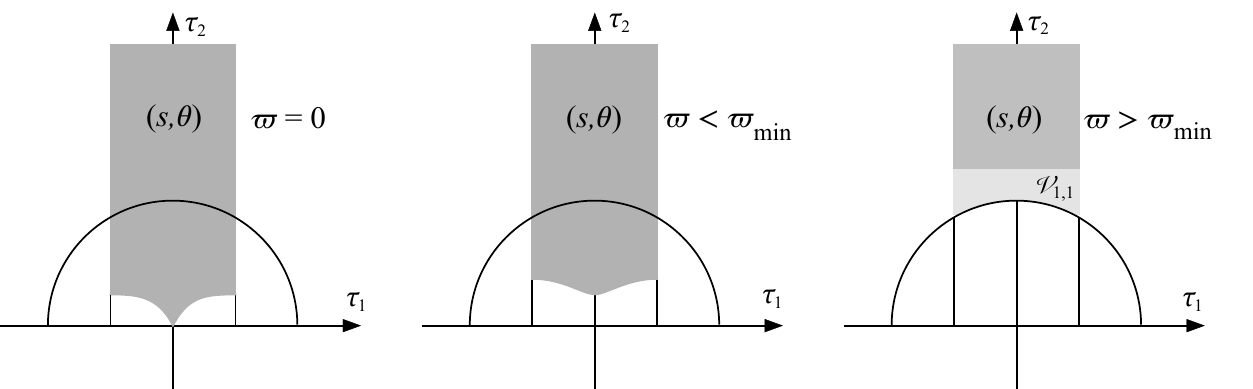}}
\end{center}
\caption{\label{fig:propcover} Region of the moduli space of the 1-punctured torus covered by the propagator diagram. When the stub length is too small, in the range $0<\varpi<\varpi_\mathrm{min}$, the covered area exceeds the fundamental region, which implies that some surfaces are counted more than once. We assume that $\varpi>\varpi_\mathrm{min}$, which leaves part of the moduli space uncovered, but does not overcount surfaces. The missing region of the moduli space defines $\mathcal{V}_{1,1}$. Note that the propagator region is bounded from below by a curve corresponding to setting the Schwinger parameter $s$ to zero. This curve attains its minimum imaginary part on the imaginary axis in the $\tau$ plane. However, for $\varpi>\varpi_\mathrm{min}$ the imaginary part of the curve is very nearly constant, as implied by \eq{tauapprox}.}
\end{figure}

Now that we have the holomorphic 1-form we can integrate to find the modulus $\tau$ as a function of $s,\theta$:
\begin{equation}
\tau = \frac{1}{2\pi i}\ln\left(\frac{2-R_-}{2-R_+}\right).\label{eq:proptau}
\end{equation}
The part of the moduli space covered by the propagator diagram depends on the stub parameter $\varpi$, as shown in figure \ref{fig:propcover}. We would like to distinguish two cases, where $\varpi$ is greater or less than a certain critical value $\varpi_\mathrm{min}$:
\begin{equation}
e^{2\varpi_\mathrm{min}} = \frac{1}{9}\Big(5+8\sinh^2\pi +4\cosh\pi\sqrt{4\sinh^2\pi +1}\Big).
\end{equation}
Numerically, the value is approximately
\begin{equation}\varpi_\mathrm{min}\approx 2.74.\end{equation}
When $\varpi<\varpi_\mathrm{min}$, the propagator diagram covers some parts of the the moduli space multiple times. While it is possible to construct a tadpole vertex that will compensate for this, this is not the standard approach and is not the one we will follow.\footnote{We thank H. Erbin for raising this point.} Instead we will assume $\varpi>\varpi_\mathrm{min}$, where the diagram leaves some region of the moduli space missing, but each surface is counted at most once. The missing region is the subset of the moduli space defining the tadpole vertex, $\mathcal{V}_{1,1}$.  Note that there is no choice of $\varpi$ such that the propagator diagram covers the whole moduli space only once. When $\varpi>\varpi_\mathrm{min}$, the modular parameter is very well approximated by the formula
\begin{equation}\tau = \frac{i}{\pi}\left(\varpi+\ln\frac{3}{2}\right)+\frac{i}{2\pi}(s-i\theta)+\mathcal{O}(e^{-2\varpi}) .\label{eq:tauapprox}\end{equation}
The complex parameter $s-i\theta$ fills a semi-infinite strip in the complex plane $\mathrm{Re}(s-i\theta)>0$ and $\mathrm{Im}(s-i\theta)\in[-\pi,\pi]$. According to \eq{tauapprox}, the part of the moduli space covered by the propagator diagram can be well approximated by scaling, rotating, and translating the rectangle $s-i\theta$ to fit inside the fundamental region of the modular parameter~$\tau$. 

\begin{figure}
\begin{center}
\resizebox{3.1in}{2in}{\includegraphics{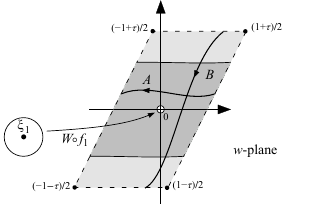}}
\end{center}
\caption{\label{fig:propunif} Picture of the propagator contribution to the tadpole diagram in the uniformization coordinate $w$. The dark grey region represents the part of the torus created by the cubic vertex, and the light grey region is the part of the torus created by the propagator tube.  The horizontal curves separating these regions is the image of the boundary of the $\xi_1$ and $\xi_\infty$ local coordinate disks. For $\varpi>\varpi_\mathrm{min}$ these curves are very nearly straight lines. The external state of the tadpole is inserted in the $\xi_1$ local coordinate, and then mapped to the origin of $w$ plane with $h_1=W\circ f_1$. For  $\varpi>\varpi_\mathrm{min}$ the image of the local coordinate is very small.}
\end{figure}

From \eq{Abel} we can transform the $z$ plane to the uniformization coordinate $w$ of the torus:
\begin{equation}
w = W(z) = \frac{1}{2\pi i}\ln\left(\frac{1-R_+}{1-R_-}\frac{z-R_-}{z-R_+}\right).\label{eq:W}
\end{equation}
This gives a new picture of the surface state as shown in figure \ref{fig:propunif}. The external state of the tadpole amplitude is inserted on the $w$-plane with the local coordinate map
\begin{equation}h_1(\xi_1) = W\circ f_1(\xi_1)=\frac{1}{2\pi i}\ln\left(\frac{1-R_+}{1-R_-}\frac{f_1(\xi_1)-R_-}{f_1(\xi_1)-R_+}\right).\label{eq:h1}\end{equation}
The local coordinate map depends on $s,\theta$ through $R_\pm$. For $\varpi\geq\varpi_\mathrm{min}$ the local coordinate map is well approximated by
\begin{equation}h_1(\xi_1) = -\frac{e^{-\varpi}}{3\pi i}\xi_1 + \mathcal{O}(e^{-3\varpi}).\label{eq:h1approx}\end{equation}
To leading order, this is simply a rescaling of $\xi_1$ which is independent of $s,\theta$. We can now write down the full contribution to the tadpole amplitude from this diagram, including $b$-ghost insertions: 
\begin{eqnarray}
-\lineup (-1)^{\Phi_1}\omega\left(\Phi_1,\ell_{0,2}\left(\frac{b_0^+}{L_0^+}\otimes\mathbb{I}\right)|\omega^{-1}\rangle\right) = -\int_0^\infty ds\int_{-\pi}^\pi \frac{d\theta}{2\pi}\Big\langle\,  h_1\circ\Phi_1(0) (W\circ f_0\circ b_0^+)\,(W\circ f_0\circ b_0^-)\Big\rangle_\tau,
\end{eqnarray}
where $b_0$ and $\overline{b}_0$ are explicitly represented as contour integrals in the $\xi_0$ coordinate,
\begin{equation}
b_0 =\oint_{|\xi_0|=1}\frac{d\xi_0}{2\pi i}\xi_0 b(\xi_0),\ \ \ \ \overline{b}_0 =\oint_{|\xi_0|=1}\frac{d\overline{\xi}_0}{2\pi i}\overline{\xi}_0 b(\overline{\xi}_0),
\end{equation}
which are transformed sequentially to the $z$ and $w$ planes with $f_0$ and $W$. The symbol $\langle\cdot\rangle_\tau$ represents the correlation function in the uniformization coordinate on the torus with modular parameter $\tau$. This completes the computation of the propagator contribution to the tadpole amplitude.

\subsection{Elementary One Loop Tadpole Vertex}
\label{subsec:Schiffer}

Now we consider the contribution to the amplitude from the tadpole vertex. The tadpole vertex is determined by the surface state
\begin{equation}
\langle \Sigma_{1,1}|\Phi_1 = \langle h_1\circ \Phi_1(0)\rangle_\tau,
\end{equation}
where the modular parameter $\tau$ lies in the region $\mathcal{V}_{1,1}$ as given in the previous subsection. The local coordinate map $h_1(\xi_1)$ is a nontrivial function of $\tau$. Since the surface state of the tadpole vertex must match that of the propagator diagram on $\d\mathcal{V}_{1,1}$, we must require
\begin{equation}h_1(\xi_1)|_{\tau\in \d \mathcal{V}_{1,1}} = W\circ f_1(\xi_1)|_{s=0}\label{eq:hcond1}\end{equation}
up to a phase rotation of the local coordinate disk. We must also require that $h_1$ varies continuously over $\mathcal{V}_{1,1}$ (up to phase rotations) and that the image of the local coordinate disk does not intersect itself in the $w$ plane.\footnote{In addition to conditions \eq{hcond1}-\eq{hcond3}, one can also require that the tadpole vertex is real \cite{Zwiebach,SenReal}. This requires that the local coordinate map satisfies
\begin{equation}h_1|_{-\tau^*} = -(h_1|_{\tau})^*.\nonumber\end{equation} The cubic vertex is real since the local coordinate maps are real.}  One might guess that we could define a suitable expression for $h_1$ by analytic continuation of the formula given in the propagator region. Unfortunately this doesn't work. Since $h_1$ must vary continuously over $\mathcal{V}_{1,1}$, it must respect the identifications (up to phase rotations):
\begin{eqnarray}
h_1(\xi_1)|_{\tau=-1/2+ib}\lineup =h_1(\xi_1)|_{\tau=1/2+ib}\label{eq:hcond2}\\
h_1(\xi_1)|_{\tau=e^{i\phi} }\lineup =\big[\tau h_1(\xi_1)\big]\big|_{\tau=e^{i(\pi-\phi)}}.\label{eq:hcond3}
\end{eqnarray}
The factor of $\tau$ in the second equation is needed since the torii at $\tau$ and $-1/\tau$ are related by scaling the $w$ plane by~$\tau$. The analytic continuation of $h_1$ from the propagator region satisfies the first identification, but not the second. Therefore we must define $h_1$ in some other way. 

Since it might seem tricky to find a definition of $h_1$ consistent with modular invariance, let us give a concrete example.\footnote{A construction of the bosonic tadpole vertex, based on the Witten cubic vertex and Strebel differentials, appears in \cite{Zemba}.} Since the local coordinate map is very closely approximated by \eq{h1approx} in the propagator region of the moduli space, it is natural to deform $h_1(\xi_1)$ through the vertex region of  the moduli space so that this approximation is exact when $|\tau|=1$:
\begin{equation}h_1(\xi_1)|_{\tau=e^{i\phi} } = -\frac{e^{-\varpi}}{3\pi i}\xi_1\label{eq:h1t1}\end{equation}
This choice does not quite satisfy the identification \eq{hcond3} at $|\tau|=1$, but it satisfies it up to a phase rotation which acts trivially on the external state of the tadpole amplitude. 
One way to implement this deformation is to replace $\varpi$ inside the local coordinate map with a nontrivial function $\widetilde{\varpi}(\tau)$ which becomes infinite at $|\tau|=1$. With a suitable normalization, the approximation \eq{h1approx} will then become exact at $|\tau|=1$. Let us describe more precisely how this works. The approximation \eq{h1approx} holds in the limit of large $\varpi$ with $s,\theta$ held fixed. For the purposes of the tadpole vertex, it is more natural to use $\tau$ as a coordinate on the moduli space. In this case, 
\eq{h1approx} holds in the limit of large $\varpi$ with
\begin{equation}\tau-\frac{i\varpi}{\pi}\end{equation}
held fixed. The local coordinate map $h_1$ in the propagator region of moduli space can be expressed as a function of $\tau$ by solving \eq{proptau} to express $R_+,R_-$ as functions of $\tau$:
\begin{eqnarray}
R_+ \lineup = 1-e^{-2\pi i\tau}-e^{-2\pi i \tau}\sqrt{1-e^{2\pi i \tau}+e^{4 \pi i \tau}}\\
R_- \lineup = 1-e^{2\pi i\tau}-\sqrt{1-e^{2\pi i \tau}+e^{4 \pi i \tau}}
\end{eqnarray}
With this, we can analytically continue the definition of $h_1$ into the tadpole vertex region of moduli space. We denote the analytic continuation as $\widetilde{h}_1$, since we reserve $h_1$ to denote the ``actual" local coordinate map of the tadpole vertex. We introduce a continuous function $\widetilde{\varpi}(\tau)$ on the tadpole vertex region of the moduli space satisfying
\begin{eqnarray}
\widetilde{\varpi}|_{\tau\in\d \mathcal{V}_{1,1}} \lineup = \varpi\label{eq:bdryprop}\\
\widetilde{\varpi}|_{\tau=-1/2+ib}\lineup = \widetilde{\varpi}|_{\tau=1/2+ib}\\
\widetilde{\varpi}|_{\tau = e^{i\phi}}\lineup =\infty
\end{eqnarray}
The local coordinate map of the tadpole vertex may then be defined:
\begin{equation}h_1(\xi_1) = e^{\widetilde{\varpi} - \varpi}\left(\widetilde{h}_1(\xi_1)\big|_{\!{\varpi\ \to\ \widetilde{\varpi} \ \ \ \ \ \ \ \ \ \ \atop\!\tau\ \to\ \tau+\frac{i}{\pi}(\widetilde{\varpi}-\varpi)}}\right)\label{eq:h1tad}\end{equation}
On the right hand side is the analytic continuation of the local coordinate map in the propagator region with the stub parameter $\varpi$ replaced with the function $\widetilde{\varpi}(\tau)$, and the modular parameter $\tau$ replaced with $\tau - i(\widetilde{\varpi}-\varpi)/\pi$. The boundary condition \eq{bdryprop} implies that $h_1$ in the tadpole vertex region matches continuously to $h_1$ in the propagator region. Towards $|\tau|=1$, the function $\widetilde{\varpi}$ becomes large while
\begin{equation}
\left(\tau+\frac{i}{\pi}(\widetilde{\varpi}-\varpi)\right)-\frac{i}{\pi}\widetilde{\varpi}
\end{equation}
remains finite. From \eq{h1approx} we then know that 
\begin{equation}
\widetilde{h}_1(\xi_1)\Big|_{{\varpi\ \to\ \widetilde{\varpi} \ \ \ \ \ \ \ \ \ \ \atop \tau\ \to\ \tau+\frac{i}{\pi}(\widetilde{\varpi}-\varpi)}}\approx -\frac{e^{-\widetilde{\varpi}}}{3\pi i}\xi_1
\end{equation}
The normalization factor in \eq{h1tad} is necessary to replace $e^{-\widetilde{\varpi}}$ in this expression---which is approaching zero---with $e^{-\varpi}$ as needed to get the desired behavior towards $|\tau|=1$. In this way we have obtained one consistent definition of the local coordinate map, but it is clear that there are many other possibilities. There is nothing about this choice which suggests it would simplify computation of higher order amplitudes. This raises the broader question as to whether there is some some principle, analogous to the minimal area problem, which extends the $SL(2,\mathbb{C})$ cubic vertex to give a natural definition of the tadpole and other higher order vertices. Presently we do not have an answer to this question.

\begin{figure}
\begin{center}
\resizebox{6in}{1.1in}{\includegraphics{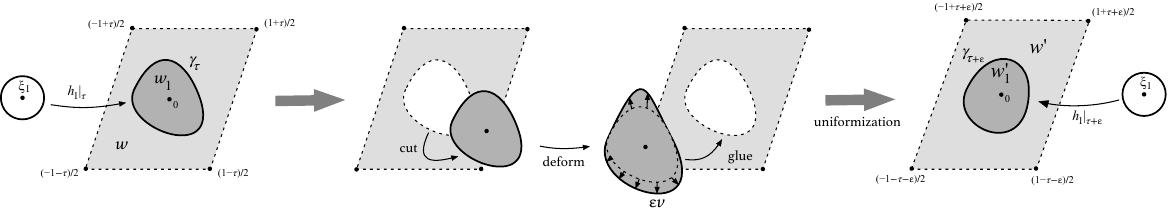}}
\end{center}
\caption{\label{fig:Schiffer} Picture illustrating the use of the Schiffer variation to deform the tadpole surface state from $\tau$ to $\tau+\eps$. The size of the local coordinate patch is exaggerated relative to figure \ref{fig:propunif} to aid visualization.}
\end{figure}

Let us assume a choice of $h_1$ has been made. The next step is to determine the $b$-ghost insertions which define the measure for integration over $\mathcal{V}_{1,1}$. Conventionally the $b$-ghosts are expressed as a contour integrals around the puncture weighted by an appropriate vector field. The vector field characterizes the Schiffer variation of the local coordinate patch, as illustrated in figure \ref{fig:Schiffer}. Consider the tadpole surface for some $\tau$ represented in the uniformization coordinate. We use $w_1$ to denote points inside the local coordinate patch, that is, points where $w_1 = h_1(\xi_1)$ for $|\xi_1|<1$ in the local coordinate disk, and we use $w$ to denote points outside. The $w$ and $w_1$ regions of the tadpole surface are separated by a curve $\gamma_\tau$. The Schiffer variation is implemented as follows. We remove the $w_1$ region from the surface, and then change its shape by deforming the boundary curve $\gamma_\tau$ along a vector field $\eps v(w_1)$ which is holomorphic in the vicinity of $\gamma_\tau$. This is the Schiffer vector field. We then glue the $w_1$ region back in such a way that the deformed boundary curve $\gamma_\tau + \eps v(\gamma_\tau)$ is identified with the original curve $\gamma_\tau$ in the tadpole surface. This procedure effectively defines a nontrivial transition function between coordinates $w$ outside the local coordinate patch and the coordinate $w_1$: 
\begin{equation}w=w_1 -\eps v(w_1).\label{eq:Schiftrans}\end{equation}
This transition function represents some deformation of the tadpole surface state. The deformation is best understood by transforming to the new uniformization coordinate. The image of the points $w,w_1$ in the new uniformization coordinate will be denoted $w',w_1'$. The Schiffer vector field is determined by two conditions. First, the modular parameter of the deformed torus must be $\tau+\eps$. Second, mapping from $\xi_1$ to $w_1$, and then from $w_1$ to $w_1'$, defines a local coordinate map from $\xi_1$ to the uniformization coordinate on the new torus. This local coordinate map is required to be identical to $h_1$ evaluated at the deformed modular parameter $\tau+\eps$.

To determine the change of the modular parameter and the new uniformization coordinate, we must compute the holomorphic 1-form. In the $w$ coordinate, this will take the general form 
\begin{equation}\omega(w)dw = (1+\eps p(w))dw.\label{eq:holw}\end{equation}
At zeroth order in $\eps$, the coordinate $w$ is identical to the uniformization coordinate and therefore $\omega(w)$ must be equal to $1$. Since $w$ satisfies the identifications $w\sim w+1$ and $w\sim w+\tau$, $p(w)$ must be a doubly periodic meromorphic function. Since the holomorphic 1-form must be holomorphic, any poles in $p(w)$ must lie in the $w_1$ region. Using the transition function \eq{Schiftrans} we may reexpress the holomorphic 1-form in the $w_1$ coordinate: 
\begin{eqnarray}
\omega(w_1)dw_1 \lineup =\omega(w)dw\nonumber\\
\lineup = \left(1+\eps \left(p(w_1) -\frac{d v(w_1)}{dw_1}\right)\right)dw_1.\label{eq:holw1}
\end{eqnarray}
The right hand side must be holomorphic in the $w_1$ region. However, $p(w_1)$ must have some singularity otherwise the holomorphic 1-form would be constant, leaving the modular parameter $\tau$ unchanged. Therefore the Schiffer vector field must have some singularity to cancel the singularity in $p(w_1)$.  It is sufficient to assume that $v(w_1)$ has a simple pole at $w_1=0$:
\begin{equation}v(w_1) = \frac{C}{w_1}+\mathrm{holomorphic},\end{equation}
with residue $C$. This implies that $p(w)$ will be a doubly periodic meromorphic function with a double pole at $w=0$. Up to a factor and additive constant, $p(w)$ must then be given by the Weierstrass $\wp$-function. It is convenient to represent the Weierstrass $\wp$-function through the second derivative of the logarithm of the theta function $\vartheta_{11}$. Then $p(w)$ will take the form
\begin{equation}
p(w) = C\frac{d^2}{dw^2}\ln\vartheta_{11}(w).
\end{equation}
The standard normalization of the holomorphic 1-form fixes the additive constant to vanish. In particular
\begin{eqnarray}
\int_A dw\,\omega(w) \lineup = 1+\eps C\int_{x}^{x+1} dw \frac{d^2}{dw^2}\ln\vartheta_{11}(w)\nonumber\\
\lineup = 1+ \eps C\left(\frac{d}{dx}\ln \vartheta_{11}(x+1)-\frac{d}{dx}\ln \vartheta_{11}(x)\right)\nonumber\\
\lineup = 1.
\end{eqnarray}
The order $\eps$ term drops out by the periodicity of $\vartheta_{11}$. Meanwhile, integrating $dw\,\omega(w)$ along the $B$ cycle determines the deformation of the modular parameter. By assumption the deformed parameter is $\tau+\eps$:
\begin{eqnarray}
\tau+\eps \lineup = \int_B dw\,\omega(w)\nonumber\\
\lineup = \tau+\eps C\left( \frac{d}{dx}\ln \vartheta_{11}(x+\tau)-\frac{d}{dx}\ln\vartheta_{11}(x)\right).
\end{eqnarray}
Using the quasi-periodicity of $\vartheta_{11}$,
\begin{equation}\vartheta_{11}(z+\tau) = e^{-2\pi i z}\vartheta_{11}(z),\end{equation}
this fixes the residue of the pole in the Schiffer vector field: 
\begin{equation}C = -\frac{1}{2\pi i}.\end{equation}
We can transform from the coordinate $w_1$ to the new uniformization coordinate $w'_1$ using the Abel map. Assuming the map preserves the origin we obtain
\begin{eqnarray}
w_1'\lineup = \int_0^{w_1} dz \left(1-\frac{\eps}{2\pi i}\frac{d^2}{dz^2} \ln\vartheta_{11}(z) -\eps\frac{d v(z)}{dz}\right)\nonumber\\
\lineup = w_1 - \frac{\eps}{2\pi i}\frac{d}{dw_1}\ln\vartheta_{11}(w_1)-\eps v(w_1).\label{eq:w1uni}
\end{eqnarray}
In dropping the boundary term at zero we have fixed the constant mode of $v(w_1)$ so that the above expression vanishes when $w_1=0$. The constant mode is not uniquely determined by our analysis since it is a globally defined holomorphic vector field which neither changes the modulus nor the local coordinate patch. The corresponding $b$ ghost insertion annihilates the tadpole surface state, and therefore does not effect the final expression for the amplitude. Therefore we are free to choose the zero mode at our convenience. 

If we identify $w_1=h_1(\xi_1)$, \eq{w1uni} defines a local coordinate map to the uniformization coordinate of deformed torus. The Schiffer vector field must be defined in such a way that this local coordinate map is identical to $h_1$ evaluated at $\tau+\eps$. There are actually two Schiffer vector fields $v_1$ and $v_2$, which deform respectively the real and imaginary parts of $\tau =\tau_1+i\tau_2$. If we deform the real part of $\tau$, the local coordinate map to the deformed torus is given by
\begin{equation} w_1' = h_1(\xi_1)|_{\tau+\eps} = h_1(\xi_1) +\eps \frac{\partial h_1(\xi_1)}{\partial \tau_1}.\end{equation}
This must be compatible with \eq{w1uni}: 
\begin{eqnarray}
w_1'= h_1(\xi_1) - \left.\frac{\eps}{2\pi i}\frac{d}{dw_1}\ln\vartheta_{11}(w_1)\right|_{w_1=h_1(\xi_1)}-\eps v_1(h_1(\xi_1)).
\end{eqnarray}
In this way we can solve for the Schiffer vector field $v_1$, and in a similar way $v_2$:
\begin{eqnarray}
v_1(w_1) \lineup = -\frac{1}{2\pi i}\frac{d}{dw_1}\ln\vartheta_{11}(w_1)-\left.\frac{\d h_1(\xi_1)}{\d\tau_1}\right|_{\xi_1=h_1^{-1}(w_1)}\label{eq:v1}\\
v_2(w_1) \lineup = -\frac{1}{2\pi}\frac{d}{dw_1}\ln\vartheta_{11}(w_1)+i\left.\frac{\d h_1(\xi_1)}{\d\tau_2}\right|_{\xi_1=h_1^{-1}(w_1)}.\label{eq:v2}
\end{eqnarray}
We have given these expressions in the uniformization coordinate of the tadpole surface with modular parameter~$\tau$. However, note that the expression
\begin{equation}\langle \Sigma_{1,1}|\frac{b(v)^2}{2!}\end{equation}
uses the Schiffer vector fields written in the local coordinate $\xi_1$ of the external state. These are related to \eq{v1} and \eq{v2} through the local coordinate map $h_1$.

With these results, the tadpole vertex can be expressed
\begin{eqnarray}
(-1)^{\Phi_1} \omega(\Phi_1,\ell_{1,0})\lineup = \frac{1}{2\pi i} \int_{\mathcal{V}_{1,1}} \langle \Sigma_{1,1}|\frac{b(v)^2}{2!}\Phi_1\\
\lineup = \frac{1}{2\pi i}\int_{\mathcal{V}_{1,1}}d\tau_1\wedge d\tau_2 \Big\langle b(v_1)b(v_2)h_1\circ \Phi_1(0)\Big\rangle_\tau,\label{eq:tadvert}
\end{eqnarray}
where
\begin{eqnarray}
b(v_1)\lineup  = \oint_{w_1=0} \frac{dw_1}{2\pi i} v_1(w_1)b(w_1) + \oint_{\overline{w}_1=0} \frac{d\overline{w}_1}{2\pi i} \overline{v}_1(\overline{w}_1)\overline{b}(\overline{w}_1)\\
b(v_2) \lineup = \oint_{w_1=0} \frac{dw_1}{2\pi i} v_2(w_1)b(w_1) + \oint_{\overline{w}_1=0} \frac{d\overline{w}_1}{2\pi i} \overline{v}_2(\overline{w}_1)\overline{b}(\overline{w}_1).
\end{eqnarray}
Finally we must fix the sign of the measure. In the propagator region the positive measure is given by $ds\wedge d\theta$, which using \eq{tauapprox} is equivalent to $d\tau_2\wedge d\tau_1$ in the $\tau$ coordinate. Since the positive measure must be the same in $\mathcal{V}_{1,1}$, \eq{tadvert} acquires a minus sign. Including the contribution from the propagator diagram, we therefore obtain 
\begin{equation}
\mathcal{A}_{1,1}^{\mathrm{bos}}(\Phi_1) = -\int_0^\infty ds\int_{-\pi}^\pi \frac{d\theta}{2\pi}\Big\langle h_1\circ \Phi_1(0)(W\circ f_0\circ b_0^+)\,(W\circ f_0\circ b_0^-)\Big\rangle_\tau-\frac{1}{2\pi i}\int_{\mathcal{V}_{1,1}}d\tau_1d\tau_2 \Big\langle b(v_1)b(v_2)h_1\circ \Phi_1(0) \Big\rangle_\tau.
\end{equation}
This completes the computation of the 1-loop tadpole amplitude in closed bosonic string field theory.

\section{One Loop Tadpole in Heterotic String Field Theory}
\label{sec:hettad} 

In this section we discuss the tadpole amplitude in heterotic string field theory. As in the bosonic string, we fix Siegel gauge $b_0^+\Phi =0$. The amplitude then takes the form
\begin{equation}
\mathcal{A}_{1,1}^{\mathrm{het}}(\Phi_1) = -(-1)^{\Phi_1}\Omega\left(\Phi_1,L_{0,2}\left(\frac{b_0^+}{L_0^+}\otimes \mathbb{I}\right)|\Omega^{-1}\rangle\right)+ (-1)^{\Phi_1}\Omega\left(\Phi_1, L_{1,0}\right).
\end{equation}
We may show that BRST exact states decouple following the same algebraic argument as presented below \eq{A10}. Noting that $\Phi_1$ must be an NS state, we may express this as 
\begin{eqnarray}
\mathcal{A}_{1,1}^{\mathrm{het}}(\Phi_1) \lineup = -(-1)^{\Phi_1}\omega\left(\Phi_1,L_{0,2}\left(\frac{b_0^+}{L_0^+}\otimes \mathbb{I}\right)|\omega^{-1}_\mathrm{NS}\rangle\right)-(-1)^{\Phi_1}\omega\left(\Phi_1,L_{0,2}\left(\frac{b_0^+X_0}{L_0^+}\otimes \mathbb{I}\right)|\omega^{-1}_\mathrm{R}\rangle\right) \nonumber\\
\lineup\ \ \ + (-1)^{\Phi_1}\omega\left(\Phi_1, L_{1,0}\right),\label{eq:hethattad}
\end{eqnarray}
where $|\omega_{\mathrm{NS}}^{-1}\rangle$ and $|\omega_{\mathrm{R}}^{-1}\rangle$ contain projections onto Neveu-Schwarz and Ramond states, respectively. To compute the amplitude we must determine $L_{0,2}$ and $L_{1,0}$. Since we already have $\ell_{0,2}$ and $\ell_{1,0}$ from the closed bosonic string, much of the work has already been done. All that remains is to insert PCOs.

\subsection{General Construction of Cubic and Tadpole Vertex}

We now give a general construction of the tree-level cubic and 1-loop tadpole vertex consistent with quantum $L_\infty$ relations. This will not completely fix the choice of PCOs. We will describe two possible choices of PCOs in the following subsections. Since we are only interested in a few simple vertices, we will not need to develop the homotopy algebra formalism and other technology of 
\cite{WittenSS,ClosedSS,Ramond} needed for constructing amplitudes to arbitrarily high order. 

In the formalism of \cite{WittenSS} and subsequent work, the data about the choice of PCOs is encapsulated in the definition of a ``contracting homotopy" for the eta zero mode acting on multi-string products. Let us review this concept. Suppose we are given an $n$-string product $b_n$ which is well-defined in the small Hilbert space. Let us also assume that $b_n$ is consistent with the closed string constraints, symmetric, and cyclic with respect to $\omega$. Since $b_n$ is well-defined in the small Hilbert space, it satisfies 
\begin{equation}[\eta,b_n] = 0\end{equation}
where we use the shorthand
\begin{equation}
[\eta,b_n] \equiv \eta b_n-(-1)^{b_n}b_n(\eta\otimes\mathbb{I}^{\otimes n-1}+...+\mathbb{I}^{\otimes n-1}\otimes \eta).
\end{equation}
Given $b_n$ we define a new product denoted $\xi\circ b_n$ satisfying 
\begin{equation}[\eta,\xi\circ b_n] = b_n\end{equation}
We assume that $\xi\circ b_n$ also satisfies closed string constraints, is symmetric and cyclic.\footnote{To be specific, $\xi\circ b_n$ is defined in the large Hilbert space, so it should be cyclic with the symplectic form $\omega$ extended in the natural way to the large Hilbert space.} We refer to $\xi\circ b_n$ as a choice ``contracting homotopy" for $\eta$ since it allows us to express $b_n$ in $\eta$-exact form. 

With this preparation, let us describe the cubic vertex. When $L_{0,2}$ multiplies two NS states at picture $-1$, it requires one PCO insertion to produce an NS state at picture $-1$. Given the bosonic product $\ell_{0,2}$, we may describe this PCO though a choice of contracting homotopy as follows: 
\begin{equation}
L_{0,2} = [Q,\xi\circ\ell_{0,2}],\ \ \ \ \ \ (\mathrm{NS\ states}).
\end{equation}
Loosely speaking, the contracting homotopy inserts $\xi$ on the product $\ell_{0,2}$, and the BRST operator turns $\xi$ into $X$. The advantage of this description is that $L_{0,2}$ is explicitly written in BRST exact form, so we automatically have
\begin{equation}[Q,L_{0,2}]=0.\end{equation}
as required by $L_\infty$ relations. In addition, the contracting homotopy is defined so that so that $L_{0,2}$ is symmetric, cyclic, and satisfies $b_0^-$ and level matching constraints.  Less obvious is that $L_{0,2}$ is meaningfully defined in the small Hilbert space. To check this, compute: 
\begin{eqnarray}
\ [\eta,L_{0,2}] \lineup = [\eta,[Q,\xi\circ\ell_{0,2}]]\nonumber\\
\lineup = -[Q,[\eta,\xi\circ\ell_{0,2}]]\nonumber\\
\lineup = -[Q,\ell_{0,2}],
\end{eqnarray}
which vanishes as a consequence of \eq{qLinf2}. This defines $L_{0,2}$ when multiplying NS states. When multiplying two Ramond states at picture $-1/2$, it must produce an NS state at picture $-1$. In this case no picture changing is required, and we may identify $L_{0,2}=\ell_{0,2}$. Following the discussion of \ref{subsec:het}, cyclicity then fixes the form of $L_{0,2}$ for both NS and R sectors:
\begin{eqnarray}
L_{0,2}=\left\{\begin{matrix}\ \ \ \ \ \ [Q,\xi\circ\ell_{0,2}], \ \ \ \ \ \ \ \ \ \ \ \ \ \ \ \  \mathrm{picture}\ (-1,-1)\Vspace\\
\ \ \ \ \ \ (\mathbb{I}+X_0)\ell_{0,2}, \ \ \ \ \ \ \ \ \ \ \ \  \ \  \, \mathrm{picture}\ (-1,-\frac{1}{2})\Vspace\\
\ \ \ \ \ \ \ \ \ \ \ \ \  \ell_{0,2},\ \ \ \ \ \ \ \ \ \ \ \ \ \ \ \  \ \ \ \ \mathrm{picture}\ (-\frac{1}{2},-\frac{1}{2})\Vspace\ 
\end{matrix}\right.,\label{eq:L02}
\end{eqnarray}
$L_{0,2}$ vanishes when multiplying any Ramond state at picture $-3/2$. This completes the definition of the cubic vertex up to a choice of contracting homotopy $\xi\circ \ell_{0,2}$.

Continuing, we may define the 1-loop tadpole vertex. The tadpole vertex is defined by a 0-string product $L_{1,0}$ subject to the relation
\begin{equation}
QL_{1,0} + L_{0,2}|\Omega^{-1}\rangle = 0.
\end{equation}
Following the general procedure of \cite{WittenSS}, we will use this equation to ``solve" for $L_{1,0}$. First note that the form of $L_{0,2}$ in this equation will depend on whether the $|\Omega^{-1}\rangle$ propagates an NS or Ramond state though the handle of the torus. We have more explicitly 
\begin{equation}
QL_{1,0} + [Q,\xi\circ\ell_{0,2}]|\omega_\mathrm{NS}^{-1}\rangle + \ell_{0,2}(X_0\otimes\mathbb{I})|\omega_\mathrm{R}^{-1}\rangle = 0.\label{eq:L10qLinf2}
\end{equation}
Next we factor $Q$ out of this equation to write
\begin{equation}
Q\left(L_{1,0} + \xi\circ\ell_{0,2}|\omega_\mathrm{NS}^{-1}\rangle -\ell_{0,2}(\xi_0\otimes\mathbb{I})|\omega_\mathrm{R}^{-1}\rangle\right) = 0.
\end{equation}
Therefore we must have 
\begin{equation}
L_{1,0} = -\xi\circ\ell_{0,2}|\omega_\mathrm{NS}^{-1}\rangle +\ell_{0,2}(\xi_0\otimes\mathbb{I})|\omega_\mathrm{R}^{-1}\rangle
+Q\lambda,
\end{equation}
where on the right hand side we add a $Q$-exact term. The $Q$-exact term is not arbitrary, but must be chosen in such a way that $L_{1,0}$ is defined in the small Hilbert space: 
\begin{equation}\eta L_{1,0}= 0.\end{equation}
Note that $\eta$ commutes through the bosonic products\footnote{The statement that $\eta$ commutes through the bosonic products amounts to the statement that we can freely deform $\eta$ contours through the corresponding surfaces. Since $\eta$ carries picture, one might worry that such contour deformations may pick up hidden residues from spurious poles. However, the sum of such residues vanishes. This follows from the fact that bosonized correlators in the large Hilbert space with some number of $\d \xi$s and only one $\xi$ are independent of the location of $\xi$. See e.g. \cite{Lechtenfeld}.} and 
\begin{equation}(\eta\otimes\mathbb{I}+\mathbb{I}\otimes \eta)|\omega^{-1}\rangle = 0.\end{equation}
Using $[\eta,\xi_0] = \mathbb{I}$ we therefore find
\begin{eqnarray}
\eta L_{1,0} \lineup = -\ell_{0,2}|\omega_\mathrm{NS}^{-1}\rangle -\ell_{0,2}|\omega_\mathrm{R}^{-1}\rangle
+\eta Q\lambda\nonumber\\
\lineup = -\ell_{0,2}|\omega^{-1}\rangle -Q(\eta\lambda).
\end{eqnarray}
This will vanish if we choose $\eta\lambda = \ell_{1,0}$ as a consequence of \eq{qLinf3}. Therefore 
\begin{equation}\lambda = \xi\circ\ell_{1,0}\end{equation}
for some choice of contracting homotopy. The 1-loop tadpole vertex will then be given by
\begin{equation}
L_{1,0} = -\xi\circ\ell_{0,2}|\omega_\mathrm{NS}^{-1}\rangle +\ell_{0,2}(\xi_0\otimes\mathbb{I})|\omega_\mathrm{R}^{-1}\rangle
+Q\big(\xi\circ\ell_{1,0}\big).\label{eq:L10Q}
\end{equation}
This completes the definition of the tadpole vertex up to a choice of contracting homotopies $\xi\circ \ell_{0,2}$ and $\xi\circ\ell_{1,0}$. 

\subsection{Case I: Local PCO insertions}
\label{subsec:Case1}

One way to give a concrete expression for the amplitude is to choose contracting homotopies which insert PCOs at specific points inside the surfaces defining the vertices. We will describe this approach in this subsection. The resulting tadpole amplitude takes a form which is broadly similar to the kind of off-shell amplitudes visualized in the work of Sen~\cite{Sen,SenWitten}.

First let us define the contracting homotopy $\xi\circ \ell_{0,2}$. Recall the picture of the bosonic cubic vertex in the global coordinate $z$, as in figure \ref{fig:cubicvert}. As a first guess, we may define $\xi\circ \ell_{0,2}$ by inserting the operator $\xi$ at some point on the $z$-plane, excluding the local coordinate patches. However, the resulting vertex will not be symmetric. To obtain a symmetric vertex, generally we will need to take a sum of six terms with $\xi$ inserted respectively at one of six points which are mapped into each other by the $S_3$ subgroup of $SL(2,\mathbb{C})$. However, in special cases $S_3$ may generate fewer than six points. The minimum number is two, which corresponds to inserting $\xi$ at $z=e^{\pm i\pi/3}$. Therefore we define
\begin{equation}
\omega(A_1,\xi\circ\ell_{0,2}(A_2,A_3))\equiv-\left\langle \left(\frac{\xi(e^{i\pi/3}) +\xi(e^{-i\pi/3})}{2}\right)f_0\circ A_1(0) f_1\circ A_2(0) f_\infty\circ A_3(0)\right\rangle.
\end{equation}
For simplicity of notation we will assume that all correlation functions are computed in the small Hilbert space. The above expression is therefore not directly meaningful. The intention, however, is that this expression should eventually appear in combinations where the $\xi$ zero mode drops out. Taking the BRST variation converts $\xi$ into $X$, and therefore the product $L_{0,2}$ of NS states is given by
\begin{equation}
\omega(A_1,L_{0,2}(A_2,A_3)) = (-1)^{A_1}\left\langle \left(\frac{X(e^{i\pi/3}) +X(e^{-i\pi/3})}{2}\right)f_0\circ A_1(0) f_1\circ A_2(0) f_\infty\circ A_3(0)\right\rangle \ \ \ \ (\mathrm{NS\ states}).
\label{eq:NScubic}\end{equation}
The product of Ramond states is already given in \eq{L02}. 

To determine the tadpole vertex we must choose a contracting homotopy $\xi\circ \ell_{1,0}$. This can be done by inserting $\xi$ at a point $p$ in the uniformization coordinate of the torus representing the bosonic tadpole vertex. Generally $p$ can be a piecewise continuous function of $\tau$ in the region $\mathcal{V}_{1,1}$, but it will be enough to assume that $p$ depends continuously on $\tau$. We require that $p$ does not encounter spurious poles or enter the local coordinate patch, but otherwise it can be chosen arbitrarily.\footnote{Reality of the tadpole vertex requires an additional condition \begin{equation}p|_{-\tau^*} = -(p|_{\tau})^*\nonumber.\end{equation} The cubic vertex is already real since the PCOs are inserted at complex conjugate locations.} The contracting homotopy is then given by
\begin{equation}
\omega(A,\xi\circ\ell_{1,0}) \equiv \sum_{\delta=\mathrm{NS,R}}\frac{1}{2\pi i}\int_{\mathcal{V}_{1,1}}d\tau_1d\tau_2\Big\langle\xi(p)b(v_1)b(v_2)h_1\circ A(0)\Big\rangle_{\tau,\delta}.
\end{equation}
Here we sum over all spin structures, which means that both NS and R states are propagating through the $A$-cycle of the torus. We label spin structures according to the corresponding theta characteristics $\delta = (a,b)$. When NS states propagate through the $A$-cycle, sum the spin structures
\begin{equation}\delta = (0,0),\ (0,1/2),\end{equation}
and likewise if Ramond states propagate through the $A$-cycle the spin structures
\begin{equation}\delta = (1/2,0),\ (1/2,1/2).\end{equation}
Having specified the contracting homotopies, we can use \eq{L10Q} to find an expression for the tadpole vertex:
\begin{eqnarray}
(-1)^{\Phi_1}\omega(\Phi_1,L_{1,0})\lineup =\sum_{\delta=\mathrm{NS}}\int_{-\pi}^\pi \frac{d\theta}{2\pi}\left.\left\langle \left(\frac{\xi\big(W(e^{i\pi/3})\big)+\xi\big(W(e^{-i\pi/3})\big)}{2}\right)(W\circ f_0\circ b_0^- )\, h_1\circ \Phi_1(0) \right\rangle_{\tau,\delta}\right|_{s=0}\nonumber\\
\lineup\ \ \ +\sum_{\delta=\mathrm{R}}\int_{-\pi}^\pi \frac{d\theta}{2\pi}\left.\Big\langle (W\circ f_0\circ\xi_0)(W\circ f_0\circ b_0^- )\,h_1\circ \Phi_1(0)\Big\rangle_{\tau,\delta}\right|_{s=0}\nonumber\\
\lineup \ \ \ -\sum_{\delta=\mathrm{NS,R}}\frac{1}{2\pi i}\int_{\mathcal{V}_{1,1}}d\tau_1d\tau_2 \Big\langle \xi(p) b(v_1)b(v_2)   \,h_1\circ Q\Phi_1(0) \Big\rangle_{\tau,\delta}.\label{eq:hettadvert}
\end{eqnarray}
Including the contribution from the propagator diagram, we obtain a complete expression for the tadpole amplitude
\begin{eqnarray}
\mathcal{A}_{1,1}^{\mathrm{het}}(\Phi_1)\lineup = -\sum_{\delta = \mathrm{NS}}\int_0^\infty ds\int_{-\pi}^\pi \frac{d\theta}{2\pi}\left\langle \left(\frac{X\big(W(e^{-i\pi/3})\big)+X\big(W(e^{-i\pi/3})\big)}{2}\right)(W\circ f_0\circ b_0^+)\,(W\circ f_0\circ b_0^-)\,h_1\circ \Phi_1(0)\right\rangle_{\tau,\delta}\nonumber\\
\lineup\ \ \ -\sum_{\delta=\mathrm{R}}\int_0^\infty ds\int_{-\pi}^\pi \frac{d\theta}{2\pi}\Big\langle (W\circ f_0\circ X_0)(W\circ f_0\circ b_0^+)\,(W\circ f_0\circ b_0^-)h_1\circ \Phi_1(0)\Big\rangle_{\tau,\delta}\nonumber\\
\lineup\ \ \ +\sum_{\delta=\mathrm{NS}}\int_{-\pi}^\pi \frac{d\theta}{2\pi}\left.\left\langle\left(\frac{\xi\big(W(e^{-i\pi/3})\big)+\xi\big(W(e^{-i\pi/3})\big)}{2}\right)(W\circ f_0\circ b_0^- )\, h_1\circ \Phi_1(0) \right\rangle_{\tau,\delta}\right|_{s=0}\nonumber\\
\lineup\ \ \ +\sum_{\delta=\mathrm{R}}\int_{-\pi}^\pi \frac{d\theta}{2\pi}\left.\Big\langle (W\circ f_0\circ\xi_0)(W\circ f_0\circ b_0^- )\,h_1\circ \Phi_1(0)\Big\rangle_{\tau,\delta}\right|_{s=0}\nonumber\\
\lineup \ \ \ -\sum_{\delta=\mathrm{NS,R}}\frac{1}{2\pi i}\int_{\mathcal{V}_{1,1}}d\tau_1d\tau_2 \Big\langle \xi(p) b(v_1)b(v_2)   \,h_1\circ Q\Phi_1(0) \Big\rangle_{\tau,\delta}.
\end{eqnarray}
At the moment it is not obvious how to evaluate these correlators in the small Hilbert space. However, since we know that the complete amplitude is independent of the $\xi$ zero mode, we may also evaluate the correlators in the large Hilbert space provided we insert an additional $\xi$. This can be achieved, for example, by replacing the vertex operator $\Phi_1(0)$ with another operator $V$ satisfying $\eta V = \Phi_1(0)$. 

However, it is desirable to express the amplitude in a form where the $\xi$ zero mode is manifestly absent. To do this we need to compute the BRST variation
\begin{equation}Q(\xi\circ\ell_{1,0}).\end{equation}
It is convenient to express this in terms of surface states:
\begin{equation}
(-1)^{\Phi_1}\omega(\Phi_1,Q(\xi\circ\ell_{1,0}))=\frac{1}{2\pi i}\int_{\mathcal{V}_{1,1}}\langle \xi\circ \Sigma_{1,1}|\frac{b(v)^2}{2!}Q\Phi_1,
\label{eq:surfQex}\end{equation}
where the state $\langle \xi\circ \Sigma_{1,1}|$ is defined for a given $\tau\in\mathcal{V}_{1,1}$
\begin{equation}
\langle \xi\circ \Sigma_{1,1}|A=\sum_{\delta = \mathrm{NS},\mathrm{R}}\Big\langle \xi(p)\, h_1\circ A(0)\Big\rangle_{\tau,\delta}.
\label{eq:xicircSig}\end{equation}
To compute the BRST variation we use
\begin{equation}\left[Q,\frac{b(v)^2}{2!}\right] = T(v)b(v)-db(v).\end{equation}
This relation, together with \eq{dmT}, implies the BRST identity \eq{BRSTid} for the bosonic tadpole vertex. Substituting into \eq{surfQex} gives
\begin{equation}
\int_{\mathcal{V}_{1,1}}\langle \xi\circ \Sigma_{1,1}|\frac{b(v)^2}{2!}Q=\int_{\mathcal{V}_{1,1}}\langle X\circ \Sigma_{1,1}|\frac{b(v)^2}{2!}-\int_{\mathcal{V}_{1,1}}\langle \xi\circ \Sigma_{1,1}|T(v)b(v)+\int_{\mathcal{V}_{1,1}}\langle \xi\circ \Sigma_{1,1}|db(v),
\label{eq:sm1}\end{equation}
where we used
\begin{equation}
\langle \xi\circ\Sigma_{1,1}|Q = \langle X\circ\Sigma_{1,1}|,
\end{equation}
and the state $\langle X\circ\Sigma_{1,1}|$ is defined by \eq{xicircSig} with $\xi$ replaced by $X$.

To proceed we must compute 
\begin{equation}\langle \xi\circ \Sigma_{1,1}|T(v).\end{equation}
If it were not for the insertion of $\xi$, this would be given by the exterior derivative of $\langle \Sigma_{1,1}|$. It is almost given by the exterior derivative of $\langle \xi\circ\Sigma_{1,1}|$, but there is an additional correction needed to ensure that the location of $\xi$ moves in the correct way after the Schiffer variation implemented by $T(v)$. Let us look at the deformation of the real part of $\tau$, implemented by the Schiffer vector field $v_1$. The energy momentum insertion deforms the surface at $\tau$ by creating a nontrivial transition function between the local coordinate patch and the region outside, as described in \eq{Schiftrans}. Passing to the new uniformization coordinate we find 
\begin{equation}
\langle \xi\circ \Sigma_{1,1}|\Big(1-\eps T(v_1)\Big)A = \sum_{\delta = \mathrm{NS},\mathrm{R}}\Big\langle\xi(p') \, h_1|_{\tau+\eps}\circ A(0)\Big\rangle_{\tau+\eps,\delta}.\label{eq:schifxi1}
\end{equation}
The Schiffer vector field $v_1$ is defined so that the torus at $\tau$ is deformed into a torus at $\tau+\eps$ (with $\eps$ is real), and the local coordinate map $h_1$ evaluated at $\tau$ has been deformed into $h_1$ evaluated at $\tau+\eps$. The Schiffer variation will move the $\xi$ insertion to a point $p'$ given by mapping $p$ to the new uniformization coordinate. Since $p$ is outside the local coordinate patch, to determine $p'$ we must first integrate the holomorphic 1-form in the $w_1$ coordinate \eq{holw1} from the puncture to a boundary point of the local coordinate patch $\hat{w}_1$; then we must integrate the holomorphic 1-form in the $w$ coordinate from the corresponding boundary point $\hat{w}$ up to the point $p$. This gives
\begin{eqnarray}
p' \lineup = \int_0^{\hat{w}_1} dw_1\, \omega(w_1) +\int_{\hat{w}}^p dw\, \omega(w)\nonumber\\
\lineup = p-\frac{\eps}{2\pi i}\frac{d}{dp}\ln\vartheta_{11}(p).
\end{eqnarray}
Therefore we can write \eq{schifxi1} more explicitly
\begin{equation}
\langle \xi\circ \Sigma_{1,1}|\Big(1-\eps T(v_1)\Big)A = \sum_{\delta = \mathrm{NS},\mathrm{R}}\left(\Big\langle\xi(p) \, h_1|_{\tau+\eps}\circ A(0)\Big\rangle_{\tau+\eps,\delta}-\frac{\eps}{2\pi i}\frac{d}{dp}\ln\vartheta_{11}(p)\Big\langle \d\xi(p) \, h_1\circ A(0)\Big\rangle_{\tau,\delta}\right).
\end{equation}
In the first term $\xi$ is inserted at the point $p$ evaluated at $\tau$. To extract a total derivative, we must replace this with $p$ evaluated at $\tau+\eps$. Therefore we rewrite
\begin{equation}
\langle \xi\circ \Sigma_{1,1}|\Big(1-\eps T(v_1)\Big)A = \sum_{\delta = \mathrm{NS},\mathrm{R}}\left(\Big\langle\xi(p|_{\tau+\eps}) \, h_1|_{\tau+\eps}\circ A(0)\Big\rangle_{\tau+\eps,\delta}-\eps\left(\frac{\d p}{\d\tau_1}+\frac{1}{2\pi i}\frac{d}{dp}\ln\vartheta_{11}(p)\right)
\Big\langle \d\xi(p) \, h_1\circ A(0)\Big\rangle_{\tau,\delta}\right).\label{eq:schifxi2}
\end{equation}
Aside from a factor of $i$ the calculation is identical for $v_2$. We therefore find
\begin{eqnarray}
\langle \xi\circ \Sigma_{1,1}| T(v_1)\lineup = -\frac{\d}{\d\tau_1}\langle\xi\circ\Sigma_{1,1}|+Z_1\langle\d\xi\circ\Sigma_{1,1}|\\
\langle \xi\circ \Sigma_{1,1}| T(v_2)\lineup = -\frac{\d}{\d\tau_2}\langle\xi\circ\Sigma_{1,1}|+Z_2\langle\d\xi\circ\Sigma_{1,1}|,
\end{eqnarray}
where the state $\langle\d\xi\circ\Sigma_{1,1}|$ is defined by \eq{xicircSig} with $\xi$ replaced by $\d\xi$, and $Z_1,Z_2$ are functions of $\tau$ defined by
\begin{eqnarray}
Z_1\lineup = \frac{\d p}{\d\tau_1}+\frac{1}{2\pi i}\frac{d}{dp}\ln\vartheta_{11}(p)\\
Z_2\lineup = \frac{\d p}{\d\tau_2}+\frac{1}{2\pi}\frac{d}{dp}\ln\vartheta_{11}(p).
\end{eqnarray}
Inserting the basis 1-forms, this becomes
\begin{equation}
\langle\xi\circ \Sigma_{1,1}|T(v) = -d\langle\xi\circ \Sigma_{1,1}| +Z\langle\d\xi\circ \Sigma_{1,1}|,
\end{equation} 
where $Z$ is the appropriate 1-form.

Substituting this result into \eq{sm1} we obtain
\begin{equation}
\int_{\mathcal{V}_{1,1}}\langle \xi\circ \Sigma_{1,1}|\frac{b(v)^2}{2!}Q=\int_{\mathcal{V}_{1,1}}\langle X\circ \Sigma_{1,1}|\frac{b(v)^2}{2!}-\int_{\mathcal{V}_{1,1}}Z\langle \d\xi\circ \Sigma_{1,1}|b(v)+\int_{\mathcal{V}_{1,1}}d\Big(\langle \xi\circ \Sigma_{1,1}|b(v)\Big).
\label{eq:sm2}\end{equation}
The last term is a total derivative on $\mathcal{V}_{1,1}$. Generally $p$ need only be a piecewise continuous function on $\mathcal{V}_{1,1}$, and integrating the total derivative will produce boundary contributions at the jumps. These are precisely the corrections due to vertical integration discussed in \cite{Sen,SenWitten}. For simplicity we will assume that $p$ is continuous on $\mathcal{V}_{1,1}$, in which case integrating the total derivative will only produce contributions from the boundary of $\mathcal{V}_{1,1}$. In particular, we assume that $p$ respects the identifications at the boundary of the fundamental region:
\begin{equation}
p|_{\tau=-1/2+ib} = p|_{\tau=1/2+ib},\ \ \ \ \ p|_{\tau=e^{i\phi}} = p|_{\tau = e^{i(\pi-\phi)}}.
\end{equation}
Then \eq{sm2} implies
\begin{eqnarray}
(-1)^{\Phi_1}\omega(\Phi_1,Q(\xi\circ\ell_{1,0}))\lineup =\sum_{\delta = \mathrm{NS,R}}\left[-\frac{1}{2\pi i}\int_{\mathcal{V}_{1,1}} d\tau_1d\tau_2\Big\langle X(p)b(v_1)b(v_2)h_1\circ\Phi_1(0)\Big\rangle_{\tau,\delta}\right.\nonumber\\
\lineup\ \ \ \ \ \ \ \ +\frac{1}{2\pi i}\int_{\mathcal{V}_{1,1}} d\tau_1d\tau_2\bigg( Z_1\Big\langle \d \xi(p)b(v_2)h_1\circ\Phi_1(0)\Big\rangle_{\tau,\delta}- Z_2\Big\langle \d \xi(p)b(v_1)h_1\circ\Phi_1(0)\Big\rangle_{\tau,\delta}\bigg)\nonumber\\
\lineup\ \ \ \ \ \ \ \ \left.-\int_{-\pi}^\pi \frac{d\theta}{2\pi}\left.\Big\langle\xi(p) (W\circ f_0\circ b_0^-)h_1\circ\Phi_1(0)\Big\rangle_{\tau,\delta}\right|_{s=0}\right].
\end{eqnarray}
In the last term we used Stokes' theorem to obtain an integral over $\d\mathcal{V}_{1,1}$, which is identical to the $s=0$ boundary of the propagator region parameterized with twist angle $\theta$. We also used \eq{bSchifbm} to exchange the $b$ ghost of the Schiffer variation with a $b_0^-$ insertion of the Poisson bivector. Substituting this result into \eq{L10Q} we obtain a new expression for the tadpole vertex:
\begin{eqnarray}
(-1)^{\Phi_1}\omega(\Phi_1,L_{1,0})\lineup =\sum_{\delta = \mathrm{NS,R}}\left[-\frac{1}{2\pi i}\int_{\mathcal{V}_{1,1}} d\tau_1d\tau_2\Big\langle X(p)b(v_1)b(v_2)h_1\circ\Phi_1(0)\Big\rangle_{\tau,\delta}\right.\nonumber\\
\lineup\ \ \ \ \ \ \ \ \left. +\frac{1}{2\pi i}\int_{\mathcal{V}_{1,1}} d\tau_1d\tau_2\bigg( Z_1\Big\langle \d \xi(p)b(v_2)h_1\circ\Phi_1(0)\Big\rangle_{\tau,\delta}- Z_2\Big\langle \d \xi(p)b(v_1)h_1\circ\Phi_1(0)\Big\rangle_{\tau,\delta}\bigg)\right]\nonumber\\
\lineup\ \ \ \ \ \ \ \ +\sum_{\delta = \mathrm{NS}}\int_{-\pi}^\pi \frac{d\theta}{2\pi}\left.\left\langle \left(\frac{\xi\big(W(e^{i\pi/3})\big)+\xi\big(W(e^{-i\pi/3})\big)}{2}-\xi(p)\right) (W\circ f_0\circ b_0^-)h_1\circ\Phi_1(0)\right\rangle_{\tau,\delta}\right|_{s=0}\nonumber\\
\lineup\ \ \ \ \ \ \ \ 
+\sum_{\delta = \mathrm{R}}\int_{-\pi}^\pi \frac{d\theta}{2\pi}\left.\left\langle \Big(W\circ f_0\circ \xi_0 - \xi(p)\Big) (W\circ f_0\circ b_0^-)h_1\circ\Phi_1(0)\right\rangle_{\tau,\delta}\right|_{s=0}.\label{eq:hettadvertsmall}
\end{eqnarray}
This is the desired expression for the tadpole vertex in the small Hilbert space. The $\xi$ zero mode is clearly absent from the first two terms. To see that is it also absent from the second two terms, we may explicitly reexpress these correlators in terms of $\d\xi$. In the third term this can be achieved by writing
\begin{equation}
\frac{\xi\big(W(e^{i\pi/3})\big)+\xi\big(W(e^{-i\pi/3})\big)}{2}-\xi(p)
= \frac{1}{2}\int_p^{W(e^{i\pi/3})} dw\, \d\xi(w) + \frac{1}{2}\int_p^{W(e^{-i\pi/3})} dw\, \d\xi(w).
\end{equation}
In the last term, we reexpress the operator $\xi_0$ as
\begin{equation}\xi_0 = \xi(x) +\Delta,\label{eq:xidecomp}\end{equation}
where $x$ is an arbitrary reference point. We can write the operator $\Delta$ in terms of $\d\xi$:
\begin{eqnarray}
\Delta \lineup =  \xi_0-\xi(x)\nonumber\\
\lineup = \oint_{|z|=1}\frac{dz}{2\pi i}\frac{1}{z}(\xi(z)-\xi(x))\nonumber\\
\lineup = \oint_{|z|=1}\frac{dz}{2\pi i}\frac{1}{z}\int_x^z dz' \,\d \xi(z').
\end{eqnarray}
Then we have
\begin{equation}W\circ f_0\circ \xi_0 - \xi(p) = W\circ f_0\circ \Delta +\int_p^{W(f_0(x))} dw\, \d\xi(w).\end{equation}
Therefore the $\xi$ zero mode disappears from the tadpole vertex, as expected. 

Note that the last two terms in \eq{hettadvertsmall} are necessary to compensate for a mismatch in the choice of PCOs between the tadpole vertex and propagator diagrams. However, this mismatch can be removed with the appropriate choice of contracting homotopy in the tadpole vertex. Suppose we instead define $\xi\circ\ell_{1,0}$ by
\begin{eqnarray}
\omega(\Phi_1,\xi\circ\ell_{1,0}) \lineup = \sum_{\delta=\mathrm{NS}}\frac{1}{2\pi i}\int_{\mathcal{V}_{1,1}}d\tau_1d\tau_2\left\langle\frac{\xi(p_1)+\xi(p_2)}{2}b(v_1)b(v_2)h_1\circ\Phi_1(0)\right\rangle_{\tau,\delta}\nonumber\\
\lineup\ \ \ +\sum_{\delta=\mathrm{R}}\frac{1}{2\pi i}\int_{\mathcal{V}_{1,1}}d\tau_1d\tau_2\left\langle\left(\int_A dw\,f(w)\xi(w)\right) b(v_1)b(v_2)h_1\circ\Phi_1(0)\right\rangle_{\tau,\delta},
\end{eqnarray}
where $p_1,p_2$ and the function $f(w)$ depend on $\tau\in\mathcal{V}_{1,1}$. In the second term $\xi$ is integrated against $f(w)$ along the $A$-cycle of the torus, and we require
\begin{equation}\int_A dw\, f(w) = 1\end{equation}
so that $\eta(\xi\circ\ell_{1,0}) = \ell_{1,0}$.  If on the boundary of $\mathcal{V}_{1,1}$ we impose the conditions
\begin{equation}\xi(p_1) = \xi\big(W(e^{i\pi/3})\big),\ \ \ \ \xi(p_2) = \xi\big(W(e^{-i\pi/3})\big),\ \ \ \ \int_A dw\,f(w)\xi(w)=W\circ f_0\circ\xi_0,\end{equation}
the PCOs will match up continuously between the tadpole vertex and propagator diagrams, and the last two terms in \eq{hettadvertsmall} will be absent. In the language of \cite{Sen}, this amplitude is obtained by integration over a weighted average of continuous sections of the fiber bundle $\widetilde{\mathcal{P}}_{1,1}$. While it is possible to choose the PCO locations continuously as a function of the moduli, one advantage of our approach is that it automatically incorporates corrections due to vertical integration when there are discontinuities in the choice of PCOs between different parts of the moduli space. 

\subsection{Spurious Poles}

We now discuss spurious poles. Our expression for the amplitude requires computing correlators of local operators in the $\xi,\eta,e^\phi$ system. Explicit formulas for such correlators were given in \cite{Verlinde}, where spurious poles are manifested by zeros of theta functions appearing in the denominator.\footnote{The expression given in \cite{Verlinde} is a correlator in the large Hilbert space. Therefore, to evaluate the amplitude using \cite{Verlinde} we must insert an additional $\xi$ at a point $x$ for each surface in the amplitude to saturate the zero mode. Since the amplitude is defined in the small Hilbert space, the final expression will be independent of $x$. For a fixed spin structure, the correlator in the large Hilbert space generally contains spurious poles at $n$ points, where $n$ is the number of insertions of $\xi$. The location of $n-1$ of these spurious poles will depend on the choice of $x$. However, since the final expression for the amplitude is independent of $x$, these spurious poles must cancel, leaving the single spurious pole described in \eq{spurious}. } We are specifically interested in correlators on the torus in the small Hilbert space, and in this case the location of spurious poles may be characterized as follows. Suppose the correlator has spin structure $\delta=(a,b)$ and contains local operators of pictures $p_1,...,p_n$ inserted respectively at points $w_1,...,w_n$ in the uniformization coordinate. A spurious pole will appear whenever the following equality holds:
\begin{equation}p_1w_1 + ....+p_n w_n = \left(\frac{1}{2}-a\right)+\tau\left(\frac{1}{2}-b\right).\label{eq:spurious}\end{equation}
For the tadpole amplitude there are two insertions that carry picture: The vertex operator $\Phi_1(0)$ at picture $-1$, inserted at the origin, and the PCO $X$ or $\xi$ at picture $+1$, inserted at some point $w$. A spurious pole will therefore appear if 
\begin{equation}w  = \left(\frac{1}{2}-a\right)+\tau\left(\frac{1}{2}-b\right).\end{equation}
We need to show that the PCOs in the tadpole amplitude do not encounter these points. The PCO configurations in the tadpole amplitude come in three kinds: 
\begin{description}
\item{(I)} Those associated with the propagator region of the moduli space with an NS internal state.
\item{(II)} Those associated with the propagator region of the moduli space with a Ramond internal state.
\item{(III)} Those associated with the tadpole vertex region of the moduli space.
\end{description}
At the interface between the propagator and tadpole vertex regions there are contributions to the amplitude with $\xi$ insertions due to vertical integration. However, since $\xi$ carries the same picture as $X$, and the location of $\xi$ corresponds directly to the location of $X$ in the vertex and propagator regions, these contributions do not need to be considered separately.
 
\begin{figure}
\begin{center}
\resizebox{6in}{1.5in}{\includegraphics{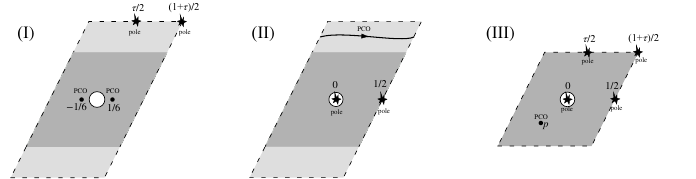}}
\end{center}
\caption{\label{fig:spurious} PCO locations versus spurious poles for the various contributions to the tadpole amplitude. Case (1) represents the propagator diagram with an NS state in the loop, case (2) represents the propagator diagram with a Ramond state in the loop, and case (3) represents the fundamental tadpole vertex.}
\end{figure}

Let us consider case (I), with an NS state in the propagator. In this part of the amplitude there is a sum of PCOs inserted respectively at 
\begin{equation}
w = W(e^{\pm i\pi/3}) = \pm\frac{1}{6}+\mathcal{O}(e^{-2\varpi}).
\end{equation}
To leading order the PCO locations are independent of $\tau$. Since we assume $\varpi\geq\varpi_\mathrm{min}$, the PCO locations will not deviate from $\pm 1/6$ by more than a few percent as a function of $\tau$. For this contribution to the amplitude we must sum over spin structures $\delta = (0,0),(0,1/2)$, and spurious poles will appear respectively at
\begin{equation}w = (1+\tau)/2,\ \tau/2.\end{equation}
Since the imaginary part of $\tau$ is never less than $1$ in the propagator region, the spurious poles are quite distant from the PCOs. Next consider case (II), with a Ramond state in the propagator. Here we have a PCO integrated along the $A$ cycle inside the propagator. For this contribution to the amplitude, we must sum over spin structures $\delta = (1/2,0),(1/2,1/2)$, and spurious poles will appear respectively at
\begin{equation}w = 0,\ 1/2.\end{equation}
The spurious poles are far away from the PCO contour in the propagator. Finally we have case (III), corresponding to the tadpole vertex. Here a single PCO appears at
\begin{equation}w = p,\end{equation}
and in the sum over all spin structures, spurious poles appear respectively at 
\begin{equation}w = 0,\ 1/2, \ \tau/2, \ (1+\tau)/2.\end{equation}
However, we have complete freedom in the choice of $p$. We may choose $p$ in such a way that it never encounters spurious poles. Therefore all contributions to the tadpole amplitude are free from singularities due to spurious poles. These results are summarized in figure \ref{fig:spurious}.

The analysis of spurious poles raises an important general question. We have defined the tadpole vertex so that it does not encounter spurious poles. But how could we have known that the propagator contribution to the amplitude does not encounter spurious poles? Of course, if it had, we would be free to make a different choice of cubic vertex. But how would we know that the redefined cubic vertex would not cause problems with other amplitudes? This is a general problem: it is not immediately clear whether a string field theory action with a fixed set of vertices will always define amplitudes that are free from singularities due to spurious poles. Sen proposed the following resolution \cite{Sen}. Assuming that:
\begin{description}
\item{(A)} The fundamental vertices do not encounter spurious poles when contracted with Fock states;
\item{(B)} The fundamental vertices come with stubs, free of operator insertions, of sufficient length;
\end{description}
amplitudes derived from the action should be free of singularities due to spurious poles. The basic intuition is that problems can only appear if the sum over intermediate states fails to converge when gluing vertices together in Feynman diagrams. If the vertices are equipped with sufficiently long stubs, the contribution from states with large conformal weight will be exponentially suppressed, and the sum over intermediate states should converge. Our calculations appear to confirm this intuition; the effect of long stubs is to freeze the PCO insertions to $\pm 1/6$ in the NS propagator contribution to the amplitude, far away from the spurious poles. However, in this particular diagram ``long" stubs are not fully necessary; even when $\varpi<\varpi_\mathrm{min}$, the PCOs from the cubic vertex do not collide with spurious poles. 

\subsection{Case II: PCO contours around punctures} 

In this subsection we describe a different approach to inserting PCOs, given by the choice of contracting homotopy
\begin{equation}
\xi\circ b_n = \frac{1}{n+1}\Big(\xi_0 b_n +(-1)^{b_n} b_n(\xi_0\otimes \mathbb{I}^{\otimes n-1}+...+\mathbb{I}^{\otimes n-1}\otimes\xi_0)\Big),\label{eq:treehom}
\end{equation}
where $\xi_0$ is the zero mode of the $\xi$ ghost:
\begin{equation}\xi_0 = \oint_{z=0}\frac{dz}{2\pi i}\frac{1}{z}\xi(z).\end{equation}
This choice was used in the construction of classical superstring field theories \cite{WittenSS,ClosedSS,Ramond}, and gives vertices characterized by contour integrals of PCOs around the punctures. The product $\xi\circ b_n$ is automatically cyclic, symmetric, and consistent with the closed string constraints provided that $b_n$ possesses these properties. The appeal of this approach is that we have a concrete formula for $\xi\circ b_n$ which does not depend on the details of the definition of $b_n$. In the classical theory, this suffices to give a unique definition of NS superstring vertices to all orders. We might hope to have similar results in the quantum theory, but spurious poles cause a complication. To fully define the vertex, we will need to carefully specify the choice of integration contours for the PCOs in relation to spurious poles. We will see why this is necessary in a moment.

In this approach, the fundamental tadpole vertex is defined by
\begin{eqnarray}
\xi\circ\ell_{0,2} \lineup = \frac{1}{3}\Big(\xi_0 \ell_{0,2} -\ell_{0,2}(\xi_0\otimes \mathbb{I} + \mathbb{I}\otimes\xi_0)\Big)\\
\xi\circ\ell_{1,0} \lineup = \xi_0\ell_{1,0},
\end{eqnarray}
and takes the form
\begin{equation}
L_{1,0} = -\frac{1}{3}\Big(\xi_0 \ell_{0,2} -\ell_{0,2}(\xi_0\otimes \mathbb{I} + \mathbb{I}\otimes\xi_0)\Big)|\omega_\mathrm{NS}^{-1}\rangle +\ell_{0,2}(\xi_0\otimes\mathbb{I})|\omega_\mathrm{R}^{-1}\rangle
+Q\big(\xi_0\ell_{1,0}\big).
\end{equation}
We may compute the BRST variation in the last term:
\begin{equation}
Q\big(\xi_0\ell_{1,0}\big) = X_0 \ell_{1,0}+\xi_0 \ell_{0,2}|\omega_\mathrm{NS}^{-1}\rangle.\label{eq:oldtadsimp}
\end{equation}
In deriving this expression we are anticipating that the $\xi_0$ contour will not need to jump across a spurious pole as a function of the modulus. If we had needed to define the $\xi_0$ contour in inequivalent ways on different regions in $\mathcal{V}_{1,1}$, we would obtain additional corrections due to vertical integration. Plugging in we therefore obtain
\begin{equation}
L_{1,0} = \left(\frac{2}{3}\xi_0 \ell_{0,2} +\frac{1}{3}\ell_{0,2}(\xi_0\otimes \mathbb{I} + \mathbb{I}\otimes\xi_0)\right)|\omega_\mathrm{NS}^{-1}\rangle +\Big(\xi_0\ell_{0,2} +\ell_{0,2}(\xi_0\otimes\mathbb{I})\Big)|\omega_\mathrm{R}^{-1}\rangle
+X_0\ell_{1,0}.
\end{equation}
Expressed in terms of correlators, the tadpole vertex is given by,
\begin{eqnarray}
(-1)^\mathrm{\Phi_1}\omega(\Phi_1,L_{1,0}) \lineup = \sum_{\delta=\mathrm{NS}}\int_{-\pi}^\pi \frac{d\theta}{2\pi}\left\langle W\circ\left(\frac{1}{3}f_0\circ\xi_0 +\frac{1}{3}f_\infty\circ\xi_0-\frac{2}{3}f_1\circ\xi_0\right)(W\circ f_0\circ b_0^-)\, h_1\circ\Phi_1(0)\right\rangle_{\tau,\delta}\nonumber\\
\lineup\ \ \ +\sum_{\delta=\mathrm{R}}\int_{-\pi}^\pi \frac{d\theta}{2\pi}\Big\langle W\circ \Big( f_0\circ\xi_0 - f_1\circ\xi_0 \Big)(W\circ f_0\circ b_0^-)\, h_1\circ\Phi_1(0)\Big\rangle_{\tau,\delta}\nonumber\\
\lineup\ \ \ -\sum_{\delta=\mathrm{NS},\mathrm{R}}\frac{1}{2\pi i}\int_{\mathcal{V}_{1,1}} d\tau_1 d\tau_2\Big\langle (W\circ f_1\circ X_0) b(v_1)b(v_2)\, h_1\circ\Phi_1(0)\Big\rangle_{\tau,\delta}.
\end{eqnarray}
Using the decomposition of $\xi_0$ in \eq{xidecomp}, it is easy to check that all correlators may be evaluated in the small Hilbert space.
Note that the contours
\begin{equation}W\circ f_0\circ\xi_0,\ \ \ \ W\circ f_\infty\circ \xi_0\end{equation}
appearing in the first term are very nearly the same due the identification $w\sim w+\tau$ on the torus. However, the contours are inequivalent due to the appearance of spurious poles at $\tau/2$ and $(1+\tau)/2$ for the NS spin structures. Deforming one contour into the other picks up residues from these spurious poles. This is related to the fact that while we have the relation 
\begin{equation}X_0\otimes\mathbb{I}|\mathrm{bpz}^{-1}\rangle = \mathbb{I}\otimes X_0|\mathrm{bpz}^{-1}\rangle,\end{equation} 
we do not have 
\begin{equation}X_0\otimes\mathbb{I}|\omega^{-1}\rangle = \mathbb{I}\otimes X_0|\omega^{-1}\rangle,\ \ \ \ (\mathrm{incorrect})\end{equation} 
due to the picture number projections contained in $|\omega^{-1}\rangle$.

\begin{figure}
\begin{center}
\resizebox{4in}{1.5in}{\includegraphics{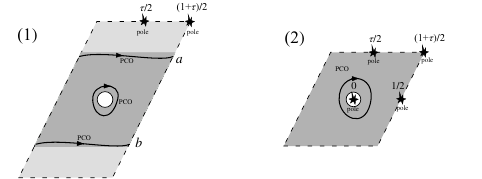}}
\end{center}
\caption{\label{fig:spuriousold} PCO locations versus spurious poles for the various contributions to the tadpole amplitude computed with the contracting homotopy \eq{treehom}. Case (1) represents the propagator diagram with an NS state in the loop. With a Ramond state in the loop, the PCO insertion is the same as in figure \ref{fig:spurious}.  Case (2) represents the fundamental tadpole vertex.}
\end{figure}

Including the contribution from the propagator diagram, the tadpole amplitude is expressed
\begin{eqnarray}
\mathcal{A}_{1,1}^{\mathrm{het}}(\Phi_1)\lineup = -\sum_{\delta = \mathrm{NS}}\int_0^\infty ds\int_{-\pi}^\pi \frac{d\theta}{2\pi}\left\langle \frac{1}{3}W\!\circ\!\Big(f_0\!\circ\! X_0 +f_\infty\!\circ \!X_0+f_1\!\circ\! X_0\Big)(W\!\circ\! f_0\!\circ\! b_0^+)\,(W\!\circ\! f_0\circ\! b_0^-)\,h_1\!\circ\! \Phi_1(0)\right\rangle_{\tau,\delta}\nonumber\\
\lineup\ \ \ -\sum_{\delta=\mathrm{R}}\int_0^\infty ds\int_{-\pi}^\pi \frac{d\theta}{2\pi}\Big\langle (W\circ f_0\circ X_0)(W\circ f_0\circ b_0^+)\,(W\circ f_0\circ b_0^-)h_1\circ \Phi_1(0)\Big\rangle_{\tau,\delta}\nonumber\\
\lineup\ \ \ +\sum_{\delta=\mathrm{NS}}\int_{-\pi}^\pi \frac{d\theta}{2\pi}\left\langle W\circ\left(\frac{1}{3}f_0\circ\xi_0 +\frac{1}{3}f_\infty\circ\xi_0-\frac{2}{3}f_1\circ\xi_0\right)(W\circ f_0\circ b_0^-)\, h_1\circ\Phi_1(0)\right\rangle_{\tau,\delta}\nonumber\\
\lineup\ \ \ +\sum_{\delta=\mathrm{R}}\int_{-\pi}^\pi \frac{d\theta}{2\pi}\Big\langle W\circ \Big( f_0\circ\xi_0 - f_1\circ\xi_0 \Big)(W\circ f_0\circ b_0^-)\, h_1\circ\Phi_1(0)\Big\rangle_{\tau,\delta}\nonumber\\
\lineup\ \ \ -\sum_{\delta=\mathrm{NS},\mathrm{R}}\frac{1}{2\pi i}\int_{\mathcal{V}_{1,1}} d\tau_1 d\tau_2\Big\langle (W\circ f_1\circ X_0) b(v_1)b(v_2)\, h_1\circ\Phi_1(0)\Big\rangle_{\tau,\delta}.
\end{eqnarray}
We may now consider spurious poles. As before, we only need to consider the contributions from NS and R propagator regions of the moduli space, and the tadpole vertex region of the moduli space. In all cases, spurious poles are avoided, as shown in figure \ref{fig:spuriousold}. In particular, spurious poles do not cross through the $X_0$ contour in the tadpole vertex region of the moduli space. This justifies the simplification of \eq{oldtadsimp}.

While we have derived a consistent expression for the tadpole amplitude, there are difficulties with an overly literal interpretation of the formula \eq{treehom} for the contracting homotopy in loop vertices. We already anticipated this when mentioning the possibility of ``anomalies" due to vertical integration in the computation of $Q(\xi_0\ell_{1,0})$. Let us explain this in more detail. The contribution to the amplitude from the tadpole vertex region of the moduli space takes the form
\begin{equation}
\omega(\Phi_1,X_0\ell_{1,0}).
\end{equation}
Since $X_0$ is BPZ even, it is tempting to interpret this as 
\begin{equation}
\omega(X_0\Phi_1,\ell_{1,0}).
\end{equation}
where $X_0\Phi_1$ is represented as a vertex operator of picture $0$. However, the amplitude expressed in this way is divergent due to a spurious pole at the puncture for the odd spin structure $\delta = (1/2,1/2)$. The point is that in this context $X_0\Phi_1$ is not really the expected picture 0 vertex operator, since the spurious pole modifies the naive result that we would obtain by computing the OPE of $X(z)$ with $\Phi_1(0)$. Another way of saying this is that $X_0\ell_{1,0}$ is not literally defined by acting the operator $X_0$ on the bosonic tadpole product $\ell_{1,0}$, since the bosonic tadpole product itself is divergent for the odd spin structure. However, in this context the interpretation of $X_0\ell_{1,0}$ is fairly clear: it is simply given by the bosonic tadpole vertex with an integration of $X(z)$ around the boundary of the local coordinate patch.

For higher order amplitudes, there are additional subtleties. In general it is not enough to interpret $\xi_0$ as integrated precisely on the boundary of the local coordinate patch, since these contours will at some point encounter spurious poles.\footnote{We encountered this issue when analyzing the 2-point amplitude in the 1PI NS heterotic string field theory.} Rather, it is necessary to allow some deformation of the $\xi_0$ contours, and these contours must be chosen in inequivalent ways in different parts of the vertex region of the moduli space to cross spurious poles. This will lead to anomalous corrections due to vertical integration when computing the BRST variation. While this is consistent, what is significant is that there is ambiguity in how the $\xi_0$ contours are specified in different regions of the moduli space. Different choices result in string field theories with different off-shell amplitudes. Therefore, the contracting homotopy \eq{treehom} will not define a unique string field theory in loops, as it does at tree level. Nevertheless, we believe that \eq{treehom}, with an appropriate interpretation of the $\xi_0$ contours, can be used to define consistent quantum superstring field theories.

\section{Concluding Remarks} 

In this paper we have given an exact computation of the off-shell 1-loop tadpole in quantum heterotic string field theory. This serves as a useful test of many recent ideas behind the construction of superstring field theories.

We focused almost exclusively on the computational problem of extracting the off-shell data that goes into the definition of the amplitude. We believe that similar analysis can be carried out for the tree-level 4-point amplitude and the 1-loop 2-point amplitude in the 1PI NS heterotic string field theory \cite{1PI}. This would be sufficient to compute, for example, the leading order contributions to the dynamically shifted vacuum state in $SO(32)$ heterotic string theory compactified on a Calabi-Yau \cite{Wittenhet,SenSUSY}.  However, the vacuum shift is not in itself an observable. To compute genuinely new physical quantities such as renormalized masses or amplitudes around the shifted vacuum, we would need higher order amplitudes. For higher orders the $SL(2,\mathbb{C})$ cubic vertex will not be enough to get exact results; we would need higher vertices with similar affinity for gluing with plumbing fixture. Probably there is a limit to how far such simplifications can extend. However, one may observe that most of the nontrivial data that emerges from the computation of the off-shell tadpole is exponentially suppressed by the stub parameter. See \eq{tauapprox} and similar equations later. This suggests that the stub parameter may be used to set up a kind of perturbation theory which could give a systematic approximation to off-shell amplitudes whose exact computation may be out of reach. This strategy may be sufficient to compute physical quantities, which in any case will be independent of the stub parameter. 

From the perspective of the formal construction of superstring field theories, our most significant conclusion is that the contracting homotopy \eq{treehom} used to insert PCOs at tree level does not generalize in a straightforward fashion to loops. In a sense this is not surprising, since there is little reason to think that a purely algebraic construction of vertices along the lines of \cite{WittenSS} can ``know" about the global conditions on Riemann surfaces that give rise to spurious poles. However, this means that the formalism of \cite{WittenSS} will not by itself provide a unique set of vertices for quantum superstring field theory. The formalism must be supplemented by data specifying a choice of contracting homotopies for each vertex so as to avoid spurious poles. It remains an interesting open question whether there is a ``best possible" choice of vertices defining the action of quantum superstring field theory to all orders.

\newpage

\noindent {\bf Acknowledgments}

\vspace{.25cm}

\noindent We would like to thank A. Sen and H. Erbin for discussions. We also thank B. Zwiebach for providing references and encouraging us to describe a concrete choice of local coordinate map in the tadpole vertex. This work is supported in part by the DFG Transregional Collaborative Research Centre TRR 33 and the DFG cluster of excellence Origin and Structure of the Universe. The work of T.E. was also supported by ERDF and M\v{S}MT (Project CoGraDS -CZ.02.1.01/0.0/0.0/15\_ 003/0000437)


\begin{thebibliography}{99}

\bibitem{Zwiebach}
 
  B.~Zwiebach,
  ``Closed string field theory: Quantum action and the B-V master equation,''
  Nucl.\ Phys.\ B {\bf 390}, 33 (1993)
  [hep-th/9206084].

\bibitem{Sonoda}

  H.~Sonoda,
  ``Hermiticity and {CPT} in String Theory,''
  Nucl.\ Phys.\ B {\bf 326}, 135 (1989).

\bibitem{Zwconst}

  B.~Zwiebach,
  ``Constraints on Covariant Theories for Closed String Fields,''
  Annals Phys.\  {\bf 186}, 111 (1988).

\bibitem{SonodaZwiebach}

  H.~Sonoda and B.~Zwiebach,
  ``Covariant Closed String Theory Cannot Be Cubic,''
  Nucl.\ Phys.\ B {\bf 336}, 185 (1990).

\bibitem{Belopolsky}

  A.~Belopolsky,
  ``Effective Tachyonic potential in closed string field theory,''
  Nucl.\ Phys.\ B {\bf 448}, 245 (1995)
  [hep-th/9412106].

\bibitem{Moeller1}

  N.~Moeller,
  ``Closed bosonic string field theory at quartic order,''
  JHEP {\bf 0411}, 018 (2004)
  [hep-th/0408067].

\bibitem{Moeller2}

  N.~Moeller,
  ``Closed Bosonic String Field Theory at Quintic Order: Five-Tachyon Contact Term and Dilaton Theorem,''
  JHEP {\bf 0703}, 043 (2007)
  [hep-th/0609209].
  
\bibitem{Pius}

  S.~F.~Moosavian and R.~Pius,
  ``Hyperbolic Geometry of Superstring Perturbation Theory,''
  arXiv:1703.10563 [hep-th].


\bibitem{WittenSS}

  T.~Erler, S.~Konopka and I.~Sachs,
  ``Resolving Witten`s superstring field theory,''
  JHEP {\bf 1404}, 150 (2014)
  [arXiv:1312.2948 [hep-th]].

\bibitem{ClosedSS}

  T.~Erler, S.~Konopka and I.~Sachs,
  ``NS-NS Sector of Closed Superstring Field Theory,''
  JHEP {\bf 1408}, 158 (2014)
  [arXiv:1403.0940 [hep-th]].
  
\bibitem{Ramond}

  T.~Erler, S.~Konopka and I.~Sachs,
  ``Ramond Equations of Motion in Superstring Field Theory,''
  JHEP {\bf 1511}, 199 (2015)
  [arXiv:1506.05774 [hep-th]].

\bibitem{complete}

  H.~Kunitomo and Y.~Okawa,
  ``Complete action for open superstring field theory,''
  PTEP {\bf 2016}, no. 2, 023B01 (2016)
  [arXiv:1508.00366 [hep-th]].


\bibitem{RWaction}

  T.~Erler, Y.~Okawa and T.~Takezaki,
  ``Complete Action for Open Superstring Field Theory with Cyclic $A_\infty$ Structure,''
  JHEP {\bf 1608}, 012 (2016)
  [arXiv:1602.02582 [hep-th]].


\bibitem{KSR}

  S.~Konopka and I.~Sachs,
  ``Open Superstring Field Theory on the Restricted Hilbert Space,''
  JHEP {\bf 1604}, 164 (2016)
  [arXiv:1602.02583 [hep-th]].

\bibitem{hetR}

  K.~Goto and H.~Kunitomo,
 ``Construction of action for heterotic string field theory including the Ramond sector,''
  JHEP {\bf 1612}, 157 (2016)
  [arXiv:1606.07194 [hep-th]].

\bibitem{Verlinde}

  E.~P.~Verlinde and H.~L.~Verlinde,
  ``Multiloop Calculations in Covariant Superstring Theory,''
  Phys.\ Lett.\ B {\bf 192}, 95 (1987).

\bibitem{SenWitten}

  A.~Sen and E.~Witten,
  ``Filling the gaps with PCO’s,''
  JHEP {\bf 1509}, 004 (2015)
  [arXiv:1504.00609 [hep-th]].
  
\bibitem{SenRev}

  C.~de Lacroix, H.~Erbin, S.~P.~Kashyap, A.~Sen and M.~Verma,
  ``Closed Superstring Field Theory and its Applications,''
  arXiv:1703.06410 [hep-th].

\bibitem{1PIR}

  A.~Sen,
  ``Gauge Invariant 1PI Effective Superstring Field Theory: Inclusion of the Ramond Sector,''
  arXiv:1501.00988 [hep-th].
  
\bibitem{SenBV}

  A.~Sen,
  ``BV Master Action for Heterotic and Type II String Field Theories,''
  JHEP {\bf 1602}, 087 (2016)
  [arXiv:1508.05387 [hep-th]].


\bibitem{Markl}

  M.~Markl,
  ``Loop homotopy algebras in closed string field theory,''
  Commun.\ Math.\ Phys.\  {\bf 221}, 367 (2001)
  [hep-th/9711045].


\bibitem{MuensterSachs}

  K.~Munster and I.~Sachs,
  ``Quantum Open-Closed Homotopy Algebra and String Field Theory,''
  Commun.\ Math.\ Phys.\  {\bf 321}, 769 (2013)
  [arXiv:1109.4101 [hep-th]].

\bibitem{fd}

  H.~Hata and B.~Zwiebach,
  ``Developing the covariant Batalin-Vilkovisky approach to string theory,''
  Annals Phys.\  {\bf 229}, 177 (1994)
  [hep-th/9301097].

\bibitem{Zwiebachmin}

  B.~Zwiebach,
  ``Quantum closed strings from minimal area,''
  Mod.\ Phys.\ Lett.\ A {\bf 5}, 2753 (1990).


\bibitem{FMS}

  D.~Friedan, E.~J.~Martinec and S.~H.~Shenker,
  ``Conformal Invariance, Supersymmetry and String Theory,''
  Nucl.\ Phys.\ B {\bf 271}, 93 (1986).
   
\bibitem{susyS}

  T.~Erler,
  ``Supersymmetry in Open Superstring Field Theory,''
  arXiv:1610.03251 [hep-th].


\bibitem{Rscatter}

  H.~Kunitomo, Y.~Okawa, H.~Sukeno and T.~Takezaki,
  ``Fermion scattering amplitudes from gauge-invariant actions for open superstring field theory,''
  arXiv:1612.00777 [hep-th].

\bibitem{susyK}

  H.~Kunitomo,
  ``Space-time supersymmetry in WZW-like open superstring field theory,''
  arXiv:1612.08508 [hep-th].
  
\bibitem{revisited}

  E.~Witten,
  ``Superstring Perturbation Theory Revisited,''
  arXiv:1209.5461 [hep-th].


\bibitem{SenSugra}

  A.~Sen,
  ``Covariant Action for Type IIB Supergravity,''
  JHEP {\bf 1607}, 017 (2016)
  [arXiv:1511.08220 [hep-th]].

\bibitem{Muenster}

  B.~Jurco and K.~Muenster,
  ``Type II Superstring Field Theory: Geometric Approach and Operadic Description,''
  JHEP {\bf 1304}, 126 (2013)
  [arXiv:1303.2323 [hep-th]].
  
\bibitem{OkawaSuperMod}

  K.~Ohmori and Y.~Okawa,
  ``Open superstring field theory based on the supermoduli space,''
  arXiv:1703.08214 [hep-th].

\bibitem{SenReal}

  A.~Sen,
  ``Reality of Superstring Field Theory Action,''
  JHEP {\bf 1611}, 014 (2016)
  [arXiv:1606.03455 [hep-th]].

\bibitem{Zemba}

  G.~Zemba and B.~Zwiebach,
  ``Tadpole Graph in Covariant Closed String Field Theory,''
  J.\ Math.\ Phys.\  {\bf 30}, 2388 (1989).

\bibitem{Sen}

  A.~Sen,
  ``Off-shell Amplitudes in Superstring Theory,''
  Fortsch.\ Phys.\  {\bf 63}, 149 (2015)
  [arXiv:1408.0571 [hep-th]].

\bibitem{Lechtenfeld}

  O.~Lechtenfeld,
  ``Superconformal Ghost Correlations On Riemann Surfaces,''
  Phys.\ Lett.\ B {\bf 232}, 193 (1989).

\bibitem{1PI}

  A.~Sen,
  ``Gauge Invariant 1PI Effective Action for Superstring Field Theory,''
  JHEP {\bf 1506}, 022 (2015)
  [arXiv:1411.7478 [hep-th]].

\bibitem{Wittenhet}

  E.~Witten,
  ``More On Superstring Perturbation Theory: An Overview Of Superstring Perturbation Theory Via Super Riemann Surfaces,''
  arXiv:1304.2832 [hep-th].


\bibitem{SenSUSY}

  A.~Sen,
  ``Supersymmetry Restoration in Superstring Perturbation Theory,''
  JHEP {\bf 1512}, 075 (2015)
  [arXiv:1508.02481 [hep-th]].




\end{thebibliography}
\end{document}